\long\def\commsingle #1\commsingleend{#1}
\long\def\commdouble #1\commdoubleend{}

\long\def\commabs #1\commabsend{#1}
\long\def\commshort #1\commshortend{#1}
\long\def\commlong #1\commlongend{}

\long\def\oldProbLowerBound #1\oldProbLowerBoundEnd{}

\long\def\commEREZDELA #1\commEREZDELAend{}

\commdouble
\documentclass{sig-alternate}

\newcommand{\redcom}[1]{{}}
\newcommand{\greencom}[1]{{}}

\commdoubleend

\commsingle
\documentclass[11pt]{article}
\usepackage{fullpage}

\usepackage{amsmath, amssymb}
\usepackage{color}
\usepackage[dvips]{graphicx}
\usepackage{epsfig,graphicx}
\usepackage[dvips]{graphicx}
\usepackage{amsmath, amssymb}
\usepackage{epsfig,graphicx}

\newcommand{\redcom}[1]{\textcolor{red} {}}
\newcommand{\greencom}[1]{\textcolor{green} {}}

\commsingleend

\newtheorem{thm}{Theorem}[section]

\newtheorem{lem}[thm]{Lemma}
\newtheorem{claim}[thm]{Claim}
\newtheorem{corollary}[thm]{Corollary}
\newtheorem{observation}[thm]{Observation}

\commsingle
\def\inline#1:{\par\vskip 7pt\noindent{\bf #1:}\hskip 10pt}
\def\midinline#1:{\par\noindent{\bf #1:}\hskip 10pt}
\def\dnsinline#1:{\par\vskip -7pt\noindent{\bf #1:}\hskip 10pt}
\def\ddnsinline#1:{\newline{\bf #1:}\hskip 10pt}

\commsingleend

\def\proof{\par\noindent{\bf Proof:~}}

\def\blackslug{\hbox{\hskip 1pt \vrule width 4pt height 8pt
    depth 1.5pt \hskip 1pt}}

\def\QED{\quad\blackslug\lower 8.5pt\null\par}

\newcommand{\blineon}[0]{\mbox{{\bf LINE}}^{\mbox{\bf on}}}
\newcommand{\rot}[0]{v_0}

\newcommand{\MCD}{{\sc MCD}}

\newcommand{\MCS}{{\sc MCD}}
\newcommand{\StRSA}{{\sc SRSA}}
\newcommand{\SRSA}{{\sc SRSA}}
\newcommand{\RSA}{{\sc RSA}}

\newcommand{\COMMIT}[0]{\mbox{\sc commit}}
\newcommand{\commit}[0]{commit}

\newcommand{\neigh}[0]{N}
\newcommand{\NI}[1]{\neigh(#1)}
\newcommand{\NRI}[1]{\neigh^{R}(#1)}
\newcommand{\NLI}[1]{\neigh^{L}(#1)}

\newcommand{\stayactive}[0]{\mbox{stays-active}}
\newcommand{\Active}[0]{\mbox{active}}

\newcommand{\Inter}[2]{{I^{#1}_{#2}}}
\newcommand{\Interk}[2]{{I^{#2}(#1)}}
\newcommand{\lfun}[1]{\ell(#1)}

\newcommand{\cost}[0]{cost}

\newcommand{\calD}{\mathcal{D}}

\newcommand{\calX}{\mathcal{X}}

\newcommand{\calL}{\mathcal{L}}

\newcommand{\calV}{\mathcal{V}}
\newcommand{\calE}{\mathcal{E}}
\newcommand{\calH}{\mathcal{H}}
\newcommand{\calA}{\mathcal{A}}
\newcommand{\calI}{\mathcal{I}}
\newcommand{\calF}{\mathcal{F}}
\newcommand{\calR}{\mathcal{R}}
\newcommand{\calQ}{\mathcal{Q}}
\newcommand{\calP}{\mathcal{P}}
\newcommand{\calC}{\mathcal{C}}

\newcommand{\naturals}{\mathbb{N}}

\newcommand{\opt}{\mbox{\textsc{opt}}}

\newcommand{\lineon}{\mbox{\sc Line}^{\mbox{\upshape on}}}
\newcommand{\lineonp}{\mbox{\sc Line}_+^{\mbox{\upshape on}}}

\newcommand{\onRSA}{\mbox{\sc srsa}^{\mbox{\upshape on}}}

\newcommand{\Base}[0]{\mbox{\sc Base}}

\newcommand{\calFT}[0]{\calF^{\mbox{\small\sc t}}}
\newcommand{\calHT}[0]{\calH^{\mbox{\small\sc t}}}
\newcommand{\calAT}[0]{\calA^{\mbox{\small\sc t}}}

\newcommand{\Aon}[0]{\calA^{\mbox{\upshape on}}}
\newcommand{\Hon}[0]{\calH^{\mbox{\upshape on}}}

\newcommand{\Fon}[0]{\calF^{\mbox{\upshape on}}}

\newcommand{\xmaxQ}[0]{\mbox{max}_x\calQ}

\newcommand{\Triangle}[0]{\mbox{\sc Triangle}}

\newcommand{\MM}[0]{M}

\newcommand{\rep}[2]{(#1,#2)}
\newcommand{\rr}{r}
\newcommand{\qq}{q}
\newcommand{\NN}{N}

\newcommand{\nn}{n}

\newcommand{\qon}{q^{\mbox{\upshape on}}}
\newcommand{\uon}{u^{\mbox{\upshape on}}}

\newcommand{\qT}{q^{\mbox{\small\sc t}}}
\newcommand{\uT}{u^{\mbox{\small\sc t}}}
\newcommand{\sT}{s^{\mbox{\small\sc t}}}

\newcommand{\rhoT}[1]{\rho^{\mbox{\small\sc t}}_{ #1}}
\newcommand{\ron}[1]{\rho^{\mbox{\upshape\small on}}_{ #1}}

\newcommand{\PHon}[0]{\calH^{\mbox{\upshape\small on}}}
\newcommand{\PVon}[0]{\calV^{\mbox{\upshape\small on}}}

\newcommand{\tminus}[2]{t^{-}_{#1}#2}
\newcommand{\tplus}[1]{t^{+}_{#1}}

\newcommand{\payer}[0]{payer}

\newcommand{\combin}[0]{\COMMIT}
\newcommand{\horbin}[0]{\calHT}
\newcommand{\arcbin}[0]{\calAT}
\newcommand{\BIN}[0]{\mbox{\sc bin}}
\newcommand{\Amin}[0]{\calA^{-1}}

\makeatletter
\newcommand{\rmnum}[1]{\romannumeral #1}
\newcommand{\Rmnum}[1]{\expandafter\@slowromancap\romannumeral #1@}
\makeatother

\begin{document}

\begin{titlepage}
\def\thepage{}

\title{
Optimal competitiveness for Symmetric Rectilinear Steiner Arborescence and related problems
\thanks{Supported in part by the Net-HD MAGNET Consortium.}
}

\author{
Erez Kantor
\thanks{Department of Electrical Engineering, Technion, Haifa, Israel.
Supported by Eshkol fellowship, the Ministry of Science and Technology, Israel.
}\\
{\small\tt erez.kantor@gmail.com}
\and
Shay Kutten
\thanks{Department of
IE\&M, Technion,
Haifa, Israel.  Supported in part by the Israel Science
Foundation and by the Technion TASP Center.
} \\
{\small\tt kutten@ie.technion.ac.il}
}

\maketitle

\begin{abstract}
We present optimal competitive algorithms for two interrelated known problems
involving Steiner Arborescence.
One is the continuous problem of the Symmetric Rectilinear Steiner Arborescence ($\SRSA$), studied by Berman and Coulston as a symmetric version of the known Rectilinear Steiner Arborescence ($RSA$) problem.

A very related, but discrete problem (studied separately in the past) is the online Multimedia Content Delivery ($\MCD$) problem on line networks, presented originally by Papadimitriu, Ramanathan, and Rangan.
An efficient content delivery was modeled as a low cost Steiner arborescence in a grid of network$\times$time they defined.
We study here the version studied by Charikar, Halperin, and Motwani (who used the same problem definitions, but removed some constraints on the inputs).

The bounds on the competitive ratios introduced separately in the above papers are similar for the two problems:  $O(\log N)$ for the continuous problem and $O(\log n)$ for the network problem, where  $N$ was the number of terminals to serve, and $n$ was the size of the network.
The lower bounds were $\Omega(\sqrt{\log N})$ and $\Omega(\sqrt{\log n})$ correspondingly. Berman and Coulston conjectured that both the upper bound and the lower bound could be improved.

We disprove this conjecture and close these quadratic gaps for both problems. We first present an $O(\sqrt{\log n})$ deterministic competitive algorithm for $\MCD$ on the line, matching the lower bound. We then translate this algorithm to become a competitive optimal algorithm $O(\sqrt{\log N})$ for $\SRSA$. Finally, we translate the latter back to solve $\MCD$ problem, this time competitive optimally even in the case that the number of requests is small
(that is, $O(\min \{\sqrt{\log n}, \sqrt{\log N}  \})$).
We also present a $\Omega(\sqrt[3]{\log n})$ lower bound on the competitiveness of any randomized algorithm. Some of the techniques may be useful in other contexts. (For example,
rather than comparing to the unknown optimum, we compared the costs of the online algorithm to the costs of an approximation {\em offline} algorithm).
\end{abstract}


\end{titlepage}

\section{Introduction}
\label{sec: Introduction}

\vspace{-0.3cm}
We present optimal online algorithms for two
 known interrelated problems involving
 Steiner Arborescences\footnote{Thee difference between a Steiner tree and a Steiner arborescence is that in the latter, directed edges are directed away from the origin.}. Those were discussed in separate studies in the past. Yet,
 we managed to improve the solution of the continuous one ($\SRSA$ \cite{berman}, defined below) by solving first the discrete one. We then used this improved solution of the continuous problem, to improve further the solution of the discrete one.
For the sake of clarity of the exposition and the motivation, let us start with the discrete problem.

 The online Multimedia Content Delivery ($\MCD$) problem on line networks  was presented originally by Papadimitriu, Ramanathan, and Rangan \cite{papa1}, to capture the tradeoff between the storage and the delivery costs. They considered a movie residing initially at some
{\em origin } node.
Requests arrived at various nodes at various times.
{\em Serving} a request meant delivering a copy to the requesting node.
An algorithm could serve every request by delivering a copy from the
origin at the time of the request, incurring a
 high delivery cost.
Alternatively, a movie already delivered to some nodes, could be stored there, and delivered later from there. This could reduce delivery costs, but incur storage cost.

More formally, given an undirected line network, Papadimitriu et al. defined a grid of network$\times$time. The full formal definitions of this this grid, the problem
and the online model, appear in Section \ref{Sec:preliminaries}. Let us now describe the ideas.
A request $r$ for a movie copy arriving at a network node $v$ at time $t$ was modeled as a request at a grid {\em vertex} $(v,t)$.
      The storage at a node $v$ from time $t$ until $t+1$ was modeled as an edge directed away from a grid vertex $(v,t)$ to vertex $(v,t+1)$. An algorithm
      had to {\em serve} the requests, in the order they arrived.
Initially, only some
origin node was served.
That is, the origin had a copy of the movie at time $0$.
Suppose that the origin continued to store this copy indefinitely. This was modeled
by an algorithm selecting the origin vertex $(0,0)$ as well as the directed path $ \{  ((0,0) , (0,1)), ((0,1) , (0,2)), ... \}$. The edge $((0,t),(0,t+1))$ modeled the storage of a copy at the
   origin $0$ from time $t$ to time $t+1$.
  Similarly, an algorithm could select some other (directed) {\em storage edges}, that is, edges of the type $((v,t),(v,t+1))$, representing the storage of a copy in $v$ from $t$ to $t+1$.
A {\em delivery} edge of the type $((v,t),(v+1,t))$
modeled the {\em delivery} of a copy, at time $t$, from network node $v$ to network node $v+1$.
As opposed to storage edges that had to be directed from $t$ to $t+1$, a delivery edge could lead either from  $(v,t)$ to $(v+1,t)$ or vice versa.

Serving a scenario (a list) of requests was modeled by an algorithm constructing a Steiner arborescence rooted at $(0,0)$
in which all the requests where terminal vertices in the grid
 network$\times$time.
 An efficient solution was a Steiner tree with a minimum number of edges (whether directed or not).
  The reader can find an example of an {\em offline} approximation  algorithm (Algorithm $\Triangle$ of Charikar et. al \cite{halperin}) in Section \ref{sec:preliminaries}.

In the online version of $\MCD$, when a request $r=(v,t)$ arrives for some network node $v$ and some time $t$, the algorithm must have already served all previous requests (those with smaller times, as well as those that have the same time but appear earlier than $r$ in the input sequence).
Moreover, the online algorithm must serve request $r$ from some vertex $(u,t)$ that is already on the solution Steiner arborescence at this point in the algorithm execution.
Hence, to be able to serve later requests, the algorithm must already add some (directed) arcs from some grid vertices of the form $(u,t)$ to the corresponding grid vertices $(u,t+1)$, since this cannot be performed later than the time all the requests for time $t$ are served.
In the case of $\MCD$ (but not of $\SRSA$), the algorithm knows when no additional requests for time $t$ will arrive, and can add such arcs $((u,t),(u,t+1))$ at that point.

The continuous version of the above problem is the Symmetric Rectilinear Steiner Arborescence ($\SRSA$) problem studied by Berman and Coulston \cite{berman} in the context of Steiner arborescences.
There, a request can arrive at any real point $(x,y)$, provided that the $Y$ coordinates are non decreasing.
Instead of selecting edges to augment the Steiner tree solution (as in $\MCD$), the algorithm may augment the Steiner arborescence by selecting either segments that is parallel to the $X$ axis, or segments that are parallel to the $Y$ axis. Papadimitriu et al. assumed some constraints on the input. Those constraints were lifted in the paper of Charikar, Halperin, and Motwani. The
 upper bound (in Charikar et al.) on the competitive ratio was
  $O(\log n)$ for the network problem (where $n$ was the size of the network)
  and the lower bound was $\Omega(\sqrt{\log n})$.
   The bounds of Berman and Coulston for $\SRSA$ were very similar. The upper bound was
  $O(\log N)$, where $N$ was the number of terminals\footnote{In fact, the parameter they used was $p$, the {\em normalized} size of the network. For simplicity, we present results for $n$, the size of the network. However, an easy consequence of our Sections \ref{sec: opt StRSA} and
\ref{sec:Optimal-mcd-for-few-requests} is that we can show the same results for $p$ rather than for $n$.}.
The lower bound was
  $\Omega(\sqrt{\log N})$. Clearly, the upper bounds are quadratic in the lower bounds.
Berman and Coulston conjectured that both the upper bound and the lower bound could be improved.

\paragraph*{Our results}
In this paper, we disprove the above conjecture and close these quadratic gaps for both problems. We first present an $O(\sqrt{\log n})$ deterministic competitive algorithm for $\MCD$ on the line.
We then translate the online algorithm to become a competitive optimal algorithm $\onRSA$
for $\SRSA$. The competitive ratio is $O(\sqrt{\log N})$.
Finally, we translate $\onRSA$ back to solve the $\MCD$ problem. This reverse translation improves the upper bound
to $O(\min \{\sqrt{\log n}, \sqrt{\log N}  \})$. That is, this final algorithm is competitive optimal for $\MCD$ even in the case that the number of requests is small. (Intuitively, the ``reverse translation'' gets rid of the dependance on the network size, using the fact that in the definition of $SRSA$, there is no network; this trick can be a useful twist on the common idea of a translation between continuous and discrete problems).

We also present a $\Omega(\sqrt[3]{\log n})$ lower bound on the competitiveness of any randomized algorithm.
Some parts of the techniques we used may be of interest. In particular, a common difficulty in computing a competitive ratio is, of course, the fact that one does not know the competing  algorithm of the adversary.
We go around this fact by comparing the costs of the online algorithm to the costs of a constant approximation offline algorithm (of Charikar, Halperin, and Motwani).

\paragraph{Some additional related work}
As pointed out in \cite{halperin}, they were also motivated by their Dynamic Servers Problem. That is, $\MCD$ is a variant of a problem that is useful for data structures for the maintenance of kinematic structures, with numerous applications. Of course,
Steiner trees, in general, have many applications, see e.g.
\cite{steiner-book} for a rather early survey that already included hundreds of items. In particular,
online Steiner {\em arborescence} problems are useful in modeling the time dimension in a process.
Intuitively, as is the case in the motivation of Papadimitriu at al. explained above,
directed edges represent the passing of time. Since there is no way to go back in time in such processes, all the directed edges are directed away from the initial state of the problem, hence, resulting in an arborescence.
Additional examples given in the literature included processes in constructing a VLSI,
optimization problems computed in iterations (where it was not feasible to return to results of earlier iterations), dynamic programming, and problems involving DNA, see, e.g. \cite{berman,CDL01,vlsi}.

Berman and Coulston also presented online algorithms for the Rectilinear
Steiner Arborescence (continuous) problem $\RSA$.
There, each horizontal line segment in the Steiner arborescence was required to be directed from a low $X$ coordinate value to a high one. (In addition, as in $\SRSA$, each vertical segment was
required to be directed from a low $Y$ coordinate value to a high one).
The offline version of $\RSA$ was studied e.g. by Rao, Sadayappan, Hwang, and Shor \cite{shor-rsa}.
$\RSA$ was attributed to \cite{natansky} who gave an exponential integer programming solution and to \cite{presented-rsa} who gave an exponential time dynamic programming algorithm.
A PTAS was presented by \cite{ptas1}.
The results of \cite{MAicalp12} generalized the logarithmic upper bound of online $\MCD$ to general networks.

\paragraph*{\bf Paper structure.}
In Section \ref{sec:MCD opt as a function of net size},
 we provide an optimal upper bound on the competitive ratio for $\MCD$ as a function of the network size.
In Section \ref{sec: opt StRSA},
we use the above solution in order to solve the (continuous) $\StRSA$ problem.
In Section \ref{sec:Optimal-mcd-for-few-requests}
we use the solution of $\StRSA$ in order to improve the solution of $\MCS$ (to be optimal also as a function of the number of Steiner points).
Finally, the lower bound is given in Section \ref{sec:randomized_lb}.

\section{Preliminaries}
\label{Sec:preliminaries}
\label{sec:preliminaries}

In this section, we present some of the definitions already given in the introduction, but in a somewhat more formal and detailed form. This allows us to introduce notations we use later.

\begin{figure}[ht]
\begin{center}
\includegraphics[width=0.4\textwidth]{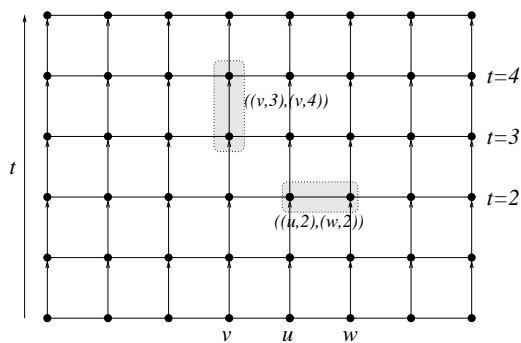}
\end{center}
\caption{\sf \label{figure:TimeNet}
An example of a time-line graph $\calL(n)=(\calV,\calE=\calH\cup\calA)$.
Each node in $\calV$ is represented by a circle;
 each horizontal edge in $\calH$ is represented by a horizontal segment (see, as an example, $(\rep{u}{2},\rep{w}{2})\in\calH$ for an horizontal edge in the left marked  rectangle);
each arc in $\calA$ is represented by a horizontal arrow (see, as an example,
$(\rep{v}{3},\rep{v}{4})\in\calA$ for an arc in the right marked  rectangle).
}
\end{figure}

\paragraph*{The network$\times$time grid}
A {\em line network } $L(\nn)=(V_\nn,E_\nn)$ is a network whose vertex set is $V_\nn=\{1,...,n\}$ and its edge set is $E_\nn=\{(i,i+1) \mid i=1,...,n-1\}$.
Given a line network $L(\nn)=(V_\nn,E_\nn)$, construct ''time-line'' graph $\calL(\nn)=(\calV_\nn,\calE_\nn=\calH_\nn\cup\calA_\nn)$, intuitively, by ``layering'' multiple copies of $V_\nn$,
one per time unit.  Connect each node in each copy to the same node in
the next copy (see Fig. \ref{figure:TimeNet}).
When it is clear from the context, we may omit $\nn$ from $X_\nn$ and write just $X$, for every $X\in\{V,E,\calV,\calH,\calA\}$.
Formally, the node set $\calV$ contains a {\em node replica}
(sometimes called just a {\em replica}) $\rep{v}{t}$ of every $v \in V$, for
every time step $t \in \naturals$.
That is, $\calV=\{\rep{v}{t} \mid v\in V, t\in \naturals \}$.
The set of edges $\calE=\calH\cup\calA$ contains \emph{horizontal edges}
$\calH=\{ (\rep{u}{t},\rep{v}{t}) \mid (u,v)\in E, t\in \naturals\}$,
connecting network edges in every time step (round),
and directed {\em vertical edges}, called {\em arcs},
$\calA=\{ (\rep{v}{t},\rep{v}{t+1}) \mid v\in V, t\in \naturals\}$,
connecting different copies of $V$.
Notice that $\calL(\nn)$ can be viewed geometrically as a square grid of $n$ by $\infty$ whose grid points are the replicas. Following Fig. \ref{figure:TimeNet}, we consider the time as if it proceeds upward.

\noindent{\bf {\em SRSA}: formal definition}
The Symmetric Rectilinear Steiner Arborescence ($\SRSA$) problem is defined as follows.
A path connecting two terminals is {\em rectilinear}
if it traverses a number of line segments, where each line segment is either vertical
or horizontal. This path is also $y$-{\em monotone} if during the traversal, the $y$ coordinates of the successive points
are never decreasing.
The input is a set of {\em requests} $\calR$, that is, a set of terminals
(sometimes called points)
$\{(x_1,y_1),...,(x_N,y_N)\}$ in the positive quadrant of the plane.
A feasible solution  $\calF$ to the problem is a set of rectilinear segments connecting all the $N$ terminals to the origin $(0,0)$ (sometime called the {\em root}) in which each
terminal can be reached from the origin by a rectilinear $y$-monotone path. The goal is to find a feasible
solution in which the sum of lengths of all the segments is the minimum possible.
%

The definition of $\MCD$ is almost identical, except that it uses $\calL(\nn)$ instead of the continuous quarter of the plane.

\noindent {\bf {\em MCD}: formal definition}
We are given a line network $L(n)$, an \emph{origin} node
 $\rot \in V$
and a set of \emph{requests} $\calR \subseteq \calV$.
%
%
A feasible solution $\calF$
is a subset of edges $\calF\subseteq\calE$ that spans the set of requests $\calR$.
For convenience, the endpoints $\calV_\calF$ of edges in $\calF$ are also considered parts of the solution.
%
%
%
For a given Algorithm $A$, let $\calF_A$ be the solution of $A$, and let $\cost(A,\calR)$, (the cost of an algorithm $A$), be $|\calF_A|$.
The goal is to find a minimum cost feasible solution.
In our analysis, $\opt$ is the set of edges in some optimal solution whose cost
is $|\opt|$.

\paragraph{\bf Online model}
\label{app:online}
In the online versions of the problems,  the algorithm receives as input a sequence of events.
One type of events is a request in the (now ordered)
set $\calR$ of requests $\calR= \{r_1, r_2, ...,r_N\}$.
A second type of events is assumed
in the case of $\MCD$ only. Specifically, we also assume for $\MCD$ a clock that tells the algorithm that time $t$ is ending, and also that time $t+1$ is starting.
This allows the algorithm (for $\MCD$ only) to know e.g. that no additional requests for time $t$ are about to arrive, or that there are no requests for some time $t$ at all.

When handling an event ${ev}$,
the algorithm only knows the following: (a) all the previous requests
$r_1, r_2, ..., r_{i}$; and (b) the solution arborescence $\calF_{ev}$ it constructed so far (originally containing only the origin). In the case of $\MCD$, it is also meaningful to say  that (c) the algorithm knows the current time $t$ (even if no request arrives at time $t$).
In each event (either a request arrival, or, in $\MCD$, a clock event), the algorithm may need to
make decisions of two types, before seeing future requests:
\begin{itemize}
\vspace{-0.2cm}
\item [(1.$\MCD$)] If the event is the arrival of a request, then from which current (time $t$) cache
(a point already in the solution arborescence $\calF_{ev}$ when $r_{i+1}$ arrives)
to serve $r_{i+1}$ by adding horizontal edges to $\calF_{ev}$.
Note that, at time $t$, the online algorithm cannot add nor delete any edge with an endpoint that corresponds to previous times.

\item [(1.$\SRSA$)] Which segments to add from a point already in the solution arborescence to
$r_{i+1}$. As opposed to the case of $\MCD$, here both horizontal and vertical segments may be added.
The segments added by the algorithm cannot include any point $(x,y)$ for $y<t_i$, where $t_i$ is the time of $r_i$.

\item[(2.$\MCD$)]
At which nodes to store a movie copy for time $t+1$, for future use.
That is, select some replica (or replicas) $(v,t)$ already in the solution
$\calF_{ev}$ and add an edge directed from $(v,t)$ to  $(v,t+1)$ to $\calF_{ev}$.

\item[(2.$\SRSA$)]
Similarly to the $\MCD$ case: first, choosing some points of the form $(x,t)$ from the points already selected to be in the solution arborescence $\calF_{ev}$ such that $t\geq t_i$; second, adding to $\calF_{ev}$ a segment directed from $(x,t)$ to some later point $(x,t^+)$.  As opposed to the case for $\MCD$, here, $t^+$ is not necessarily $t+1$, so that algorithm also must choose $t^+$.

\end{itemize}

\noindent
Similarly to \cite{ABF93,MAicalp12,BFR92,papa1,papa2,papa3,halperin},
we assume that the online algorithm may replicate the movie for
efficient delivery, but at least one copy of the movie must remain in
the network at all times.  Alternatively, the system (but not the
algorithm) can have the option to delete the movie altogether, this
decision is then made known to the online algorithm.
This natural assumption is also necessary for having a competitive algorithm.

\begin{figure*}
\begin{center}
\includegraphics[width=0.6\textwidth]{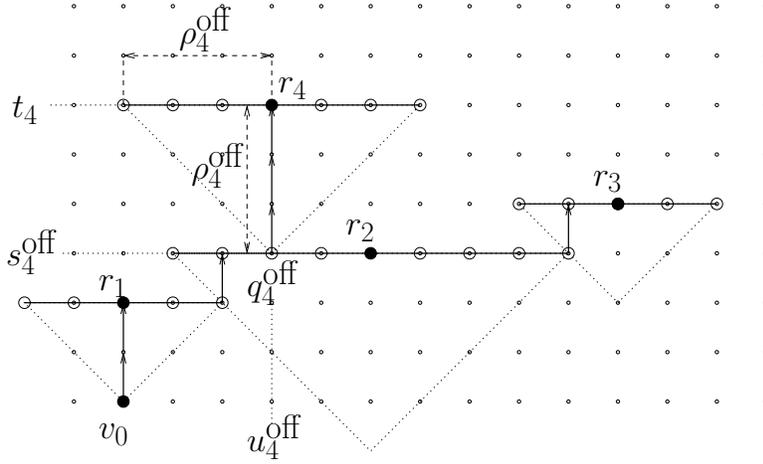}
\end{center}
\caption{\sf \label{figure:VerPathTriangle}\label{figure: Triangle}
An example of an execution of $\Triangle$ for requests set $\calR=\{r_1,r_2,r_3,r_4\}$.
The non-filled circles correspond to base replicas.
The set of horizontal edges $\calHT$ is the union of the bases of the triangles;
The set of arcs $\calAT$ is the union of the vertical paths from the serving replicas $\qT_i=(\uT_i,\sT_i)$ to the base replica $\rep{\uT_i}{t_i}$ (for $i=1,...,4$).}
\end{figure*}

\paragraph*{\bf A tool: the offline algorithm $\Triangle$ of Charikar et. al}
Consider a requests set $\calR=\{\rr_0=\rep{\rot}{0},\rr_1=\rep{v_1}{t_1},...,\rr_\NN=\rep{v_\NN}{t_\NN} \}$ such that $0\leq t_1\leq t_2\leq...\leq t_\NN$.
When Algorithm $\Triangle$ starts, the solution includes just $\rr_0=\rep{\rot}{0}$
(intuitively, a ``pseudo request'').
Then, $\Triangle$ handles, first, request $\rr_1$, then request $\rr_2$, etc...
In handling a request $\rr_i$, the algorithm may add some edges to the solution.
(It never deletes any edge from the solution.)
After handling $\rr_i$, the solution is an arborescence rooted at $\rr_0$ that spans the request replicas $\rr_1,...,\rr_i$.
For each such request $\rr_i\in\calR$, $\Triangle$ performs the following (see Fig. \ref{figure: Triangle}).

\begin{itemize}
\item [(T1)] Chose a replica $\qT_i=\rep{\uT_i}{\sT_i}$ s.t. $\qT_i$ is already in the solution and the distance from $\qT_i$ to $\rr_i$ is minimum (over the replicas already in the solution). Call $\qT_i$ the {\em serving replica} of $r_i$.

\item [(T2)] Define the {\em radius} $\rhoT{i}$ of $\rr_i$ as $\rhoT{i}=d(\qT_i,\rr_i)$.
Also define the {\em base}\footnote{The word ``base'' comes from the notation used in \cite{halperin} for Algorithm $\Triangle$. There, $\Base(i)$ is the base of the triangle defined there (that triangle is illustrated in Fig. \ref{figure: Triangle}).
} $\Base(i)$ of $\rr_i$ as the set of replicas at time $t$ of distance at most $\rhoT{i}$ from $\rr_i$.
That is, $\Base(i)=\{q=(v,t)\in\calV \mid d(\rr_i,q)\leq\rhoT{i} \mbox{ and } t_i=t\}$.
Similarly, the {\em edge base} of $\rr_i$ is $\Base_\calH(i)=\{(\rr,\qq)\in\calH \mid \rr,\qq\in\Base(i)\}$.

\item[(T3)] Deliver a copy to a replica in $\Base(i)$.
This is done by delivering a copy from $\rep{\uT_i}{\sT_i}$ to $\rep{\uT_i}{t_i}$ (meaning that node $\uT_i$ stores a copy from time $\sT_i$ to time $t_i$).
More formally, add the arcs of $\calP_\calA[\rep{\uT_i}{\sT_i},\rep{\uT_i}{t_i}]$ to the solution.

\item[(T4)] Deliver a copy to all replicas in $\Base(i)$.
This is done by adding
all the edges of $\Base_\calH(i)$ to the solution, except the one that closes a circle\footnote{
For convenience, of the analysis we want the solution to be a tree, so we do not add redundant edge.
} (if such exists).

\end{itemize}

It is easy to verify \cite{halperin} that the cost of $\Triangle$ for serving
the $i$'th request $\rr_i$ is $3\rhoT{i}$ at most.
Denote by $\calFT=\calHT\cup\calAT$ the feasible solution of $\Triangle$, where
$\calHT\subseteq\cup_{i=1}^{\NN}\Base_\calH(i)$ and $\calAT=\cup_{i=1}^{\NN}\calP_\calA[\rep{\uT_i}{\sT_i},\rep{\uT_i}{t_i}]$.
Note that $\calFT$ is an arborescence rooted at $\rep{\rot}{0}$ spanning the base replicas of $\Base=\cup_{i=1}^{\NN}\Base(i)$.
Rewording the theorem of \cite{halperin}, somewhat,
\begin{thm}
\cite{halperin}
$\Triangle$ computes a $3$-approximate solution. Also, $\sum_{i=1}^{N}\rhoT{i}\leq |\opt|$.
\label{thm: Triangle is a 3-approx}
\end{thm}

\begin{figure}[ht]
\begin{center}
\includegraphics[width=0.4\textwidth]{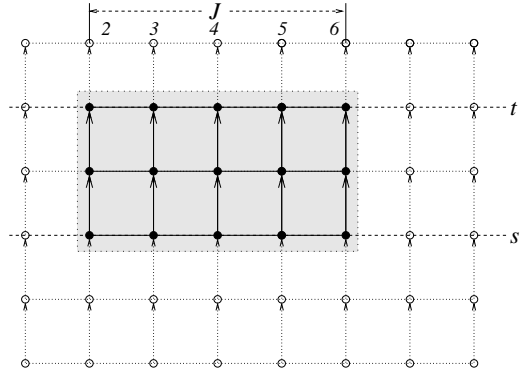}
\end{center}
\caption{\sf \label{figure:subgraphJ}
A subgraph rectangle $J[s,t]$, where $J=\{2,3,4,5,6\}$.}
\end{figure}

\paragraph*{\bf General definitions and notations.}
Consider an interval $J=\{v,v+1,...,v+\rho\}\subseteq V$
and two integers $s,t\in\naturals$, s.t. $s\leq t$.
Let $J[s,t]$ (Fig. \ref{figure:subgraphJ}) be the
{\em ``rectangle subgraph''} of $\calL(\nn)$ corresponding to vertex set $J$ and time interval $[s,t]$.
This rectangle consists of the
replicas and edges of the nodes of $J$ corresponding to time interval $[s,t]$.
For a given subsets $\calV'\subseteq \calV$, $\calH'\subseteq\calH$ and $\calA'\subseteq\calA$,
denote by
(1) $\calV'[s,t]$ replicas of $\calV'$ corresponding to times
 $s,...,t$. Define similarly (2) $\calH'[s,t]$ for horizontal edges of $\calH'$; and (3) $\calA'[s,t]$ arcs of $\calA'$.
(When $s=t$, we may write
$\calX[t]=\calX[s,t]$, for $\calX\in\{J,\calV',\calH'\}$.)

Consider also two nodes $v,u\in V$ s.t. $v\leq u$.
Let $\calP_\calH[\rep{v}{t},\rep{u}{t}]=\calP_\calH[\rep{u}{t},\rep{v}{t}]$
be the set of horizontal edges of the shortest path from $\rep{v}{t}$ to $\rep{u}{t}$.
That is, $\calP_\calH[\rep{v}{t},\rep{u}{t}]=\{(\rep{w}{t},\rep{(w+1)}{t})\mid v\leq w<u\}$.
Let $\calP_\calA[\rep{v}{s},\rep{v}{t}]$ be the set of arcs of the shortest path from $\rep{v}{s}$ to $\rep{v}{t}$.
That is, $\calP_\calA[\rep{v}{s},\rep{v}{t}]=\{(\rep{v}{z},\rep{v}{z+1})\mid s\leq z<t\}$.
Let $d(\rep{u}{s},\rep{v}{t})$ be the distance from $\rep{u}{s}$ to $\rep{v}{t}$.
Formally, $d(\rep{u}{s},\rep{v}{t})=t-s+|v-u|$ (if $s\leq t$, otherwise, 
$\infty$).

\section{{\large\bf Optimal online algorithm for MCD}
\commdouble\\ \commdoubleend
}
\label{sec:MCD opt as a function of net size}

\paragraph*{\bf Algorithm $\blineon$.}
\label{subsec: online on the line sqrt(log n)}
Like Algorithm $\Triangle$, Algorithm $\lineon$, handles requests one by one, according to the order of arrival.
However, in step (T3), $\Triangle$ may perform an operation that no online algorithm can perform (if $\sT_i<t_i$).
Serving a request $\rr_i$ must be preformed from some replica $\qon_i=\rep{\uon_i}{t_i}\in\calV[t_i]$ that holds a copy at time $t_i$ in the execution of the online algorithm on $\calR$.
Thus (in addition to selecting from which nodes to deliver copies), algorithm
$\lineon$ at time $t_i-1$ had to also select the nodes that store copies for the consecutive time $t_i$ (so that $\qon_i$ mentioned above would be one of them).
Let us start with some definitions.

\begin{figure}[ht!]
\begin{center}
\includegraphics[width=0.45\textwidth]{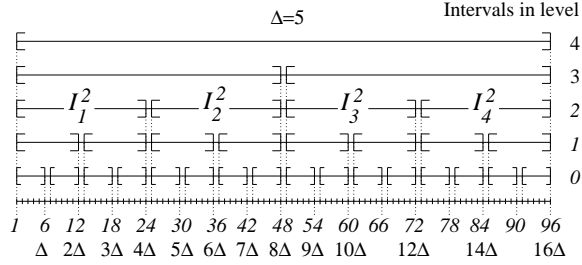}
\end{center}
\caption{\sf \label{figure:IntervalsA} An example of a line network of size $n=16\Delta=96$, where $\Delta=6$. There are $\log (96/\Delta) +1=5$ partition levels. At level $l$, the interval [1,96] partitions into $2^{4-l}=96/\Delta2^l$ intervals $\Inter{l}{1},...,\Inter{l}{2^{4-l}}$, each of size $\Delta2^l$ (see for example, level 3).
}
\end{figure}

\begin{figure*}
\begin{center}
\includegraphics[width=0.45\textwidth]{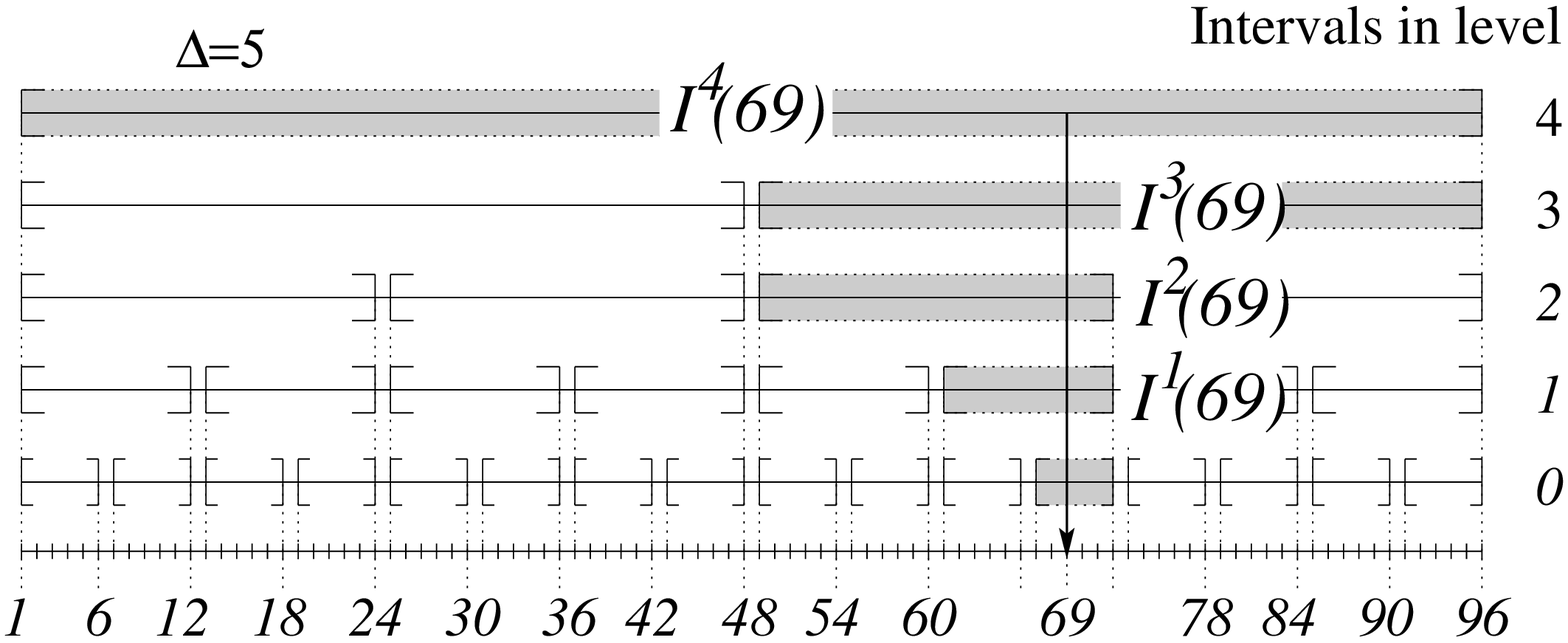}
\end{center}
\caption{\sf \label{figure:IntervalsB}
An example of the membership of node 69 in intervals in a network of size 96.
Node 69 belongs to 5 intervals, $\Interk{69}{0}=\Inter{0}{12},~ \Interk{69}{1}=\Inter{1}{6},~ \Interk{69}{2}=\Inter{2}{3},~ \Interk{69}{3}=\Inter{3}{2}$ and $\Interk{69}{4}=\Inter{4}{1}$.
}
\end{figure*}

\paragraph*{\bf Partitions of $[1,n]$ into intervals.}
Define $m=n/\Delta$ for some positive integer $\Delta$ to be chosen later.
For convenience, we assume that $m=n/\Delta$ is a power of $2$.
(It is trivial to generalize it).
Define $\log m+1$ {\em levels} of partitions of the interval $[1,n]$.
In level $l$, partition $[1,n]$ into $m/2^l=n/\Delta2^l$ intervals,
$\Inter{l}{1}$, $\Inter{l}{2}$,...,$\Inter{l}{m/2^l}$, each of size $\Delta2^l$
(Fig. \ref{figure:IntervalsA}).
$\Inter{l}{j}=\{\Delta(j-1)\cdot 2^l+k\mid k=1,...,\Delta2^l\}$,
for every $1\leq j \leq m/2^{l}$ and every $0\leq l \leq \log m$.
Let $\calI$ be the set of all such intervals.
Let $\lfun{I}$ be the {\em level} of an interval $I\in\calI$, i.e., $\lfun{\Inter{l}{j}}=l$.
Denote by $\Interk{v}{l}$ (for every node $v\in V$ and every level $l=0,..., \log m$) the  interval in level $l$ that contains $v$.
That is,
$\Interk{v}{l}=\Inter{l}{k}, \mbox{ where } k=\left\lfloor\frac{v}{\Delta2^l}\right\rfloor+1,$
(Fig. \ref{figure:IntervalsB}).

\begin{figure*}
\begin{center}
\includegraphics[width=0.4\textwidth]{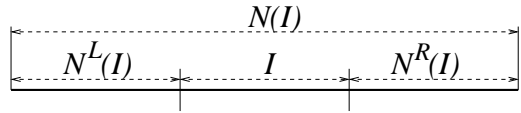}
\end{center}
\caption{\sf \label{figure:NeighborsA}
Neighbor intervals.
}
\end{figure*}

For a given interval $\Inter{l}{j}\in\calI$,
denote by $\NRI{\Inter{l}{j}}$, for $1\leq j<m/2^l$ (respectively, $\NLI{\Inter{l}{j}}$, for $1< j\leq m/2^l$) the {\em neighbor} interval of level $l$ that is on the right (resp., left) of $\Inter{l}{j}$ (see Fig. \ref{figure:NeighborsA}).
That is, $\NLI{\Inter{l}{j}}=\Inter{l}{j-1}$ and $\NRI{\Inter{l}{j}}=\Inter{l}{j+1}$.
Define that $\NLI{\Inter{i}{1}}=\emptyset$ and $\NRI{\Inter{i}{m/2^l}}=\emptyset$.
Let
$$\NI{\Inter{}{}}=
\NLI{\Inter{}{}}\cup \Inter{}{}\cup\NRI{\Inter{}{}}.
$$
We say that $\NI{I}$ is the {\em neighborhood } of $I$.

\begin{figure*}
\begin{center}
\includegraphics[width=0.4\textwidth]{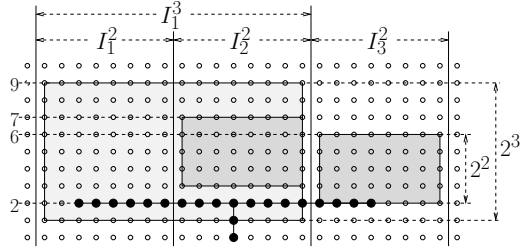}
\end{center}
\caption{\sf \label{figure:DeltaActive}
Interval $I_1^3$ is $\stayactive$ at time $t=9$; interval $I^2_2$ is {\em not} active at time $t=7$;
and interval $I_3^2$ is active, but {\em not} $\stayactive$ at time $t=6$.
}
\end{figure*}

\paragraph*{\bf Active intervals.}
\label{subsection: def delta active intervals}
An interval $\Inter{}{}\in\calI$ is called {\em active} at time $t$,
if a replica in $\Inter{}{}[t-2^{\lfun{I}},t]$ is also in $\Base$, i.e., $\Base\cap I[t-2^{\lfun{I}},t]\not=\emptyset$
(see Fig. \ref{figure:DeltaActive}).
Intuitively, the pseudo online kept a movie copy in, at least, one of the nodes of $I$, at least once, and ``not to long'' before time $t$.
We say that $I$ {\em $\stayactive$}, intuitively, if $I$ is {\bf not} ``just about to stop being active'',
that is, if $\Base\cap I[t-2^{\lfun{I}}+1,t]\not=\emptyset$.

Denote by $\calC_{t+1}$, the set of replicas corresponding to the nodes that store copies from time $t$ to time $t+1$ in a $\lineon$ execution.
Also, $\calC_0=\{\rr_0=\rep{\rot}{0}\}$.
We chose to leave a copy in $\rot$ always.
To help us later in the analysis, we also added an auxiliary set $\COMMIT\subseteq\{\langle I,t\rangle\mid I\in\calI,t\in\mathbb{N}\}$.
Initially, $\COMMIT\leftarrow\emptyset$.
For each time $t=0,1,2,...$, consider first the case that there exists at least one request corresponding to time $t$, i.e., $\calR[t]=\{\rr_{j},...,\rr_k\}\not=\emptyset$.
Then, for each request $\rr_i\in\calR[t]$,
$\lineon$ simulates $\Triangle$ to find the radius $\rhoT{i}$ and the set of base replicas $\Base(i)$ of $\rr_i$.
Next, $\lineon$ delivers a copy to every such base replica $\rr\in\Base(i)$ (this is called the {\em ``delivery phase''}).
That is, for each $i=j,...,k$ do:

\begin{itemize}

\item[(D1)] chose a closest (to $\rr_i$) replica $\qon_i=(\uon_i,t)$ of time $t$ already in the solution;

\item[(D2)] add the path $\PHon(i)=\calP_\calH[\qon_i,\rr_i]\cup\Base_\calH(i)$ to the solution.

\end{itemize}

Let $\PVon(i)=\{\rr \mid (r,q)\in\PHon(i)\}$. (Note that $\rr_j$ is served from $\calC_t$, after that, the path $\PHon(j)$ is added; and $\rr_{j+1}$ is served from $\calC_t\cup\PVon(j)$, etc.)
Clearly, the delivery phase of time $t$ ensures that
(at least) the nodes of $\calC_{t}\cup\Base[t]$ have copies at the end of that phase.
It is left to decide which of the above copies to leave for time $t+1$.
That is (the {\em ``storage phase''}),
$\lineon$ chooses the set $\calC_{t+1}\subseteq \calC_{t}\cup\Base[t]$.
Initially, $C_{t+1}\leftarrow\{\rep{\rot}{t+1}\}$ (as we chose to leave a copy at $\rot$ always).
Then, for each level $l=0,...,\log m$ in an {\em increasing} order select as follows.

\begin{itemize}

\item[(S1)] While there exists a level $l$ interval $I\in\calI$ that is ($\rmnum{1}$) $\stayactive$ at $t$;
but ($\rmnum{2}$) no replica has been selected in $I$'s neighborhood (i.e., $\calC_{t+1}\cap\NI{I}[t+1]=\emptyset$),
then perform steps (S1.1-S1.3) below.

\item[(S1.1)] Add the tuple $\langle I,t\rangle$ to the 
set $\COMMIT$ (we say that $I$ {\em commits} at time $t$).

\item[(S1.2)] Select some replica $\rep{v}{t}\in \Base[t]\cup \calC_{t}$ such that $v\in\NI{I}$
(by Observation \ref{obser: complete onalg} below, such a replica does exist).

\item[(S1.3)] Add  $\rep{v}{t+1}$ to $\calC_{t+1}$ and add the arc $(\rep{v}{t},\rep{v}{t+1})$ to the solution.

\end{itemize}

\begin{figure*}
\begin{center}
\includegraphics[width=0.4\textwidth]{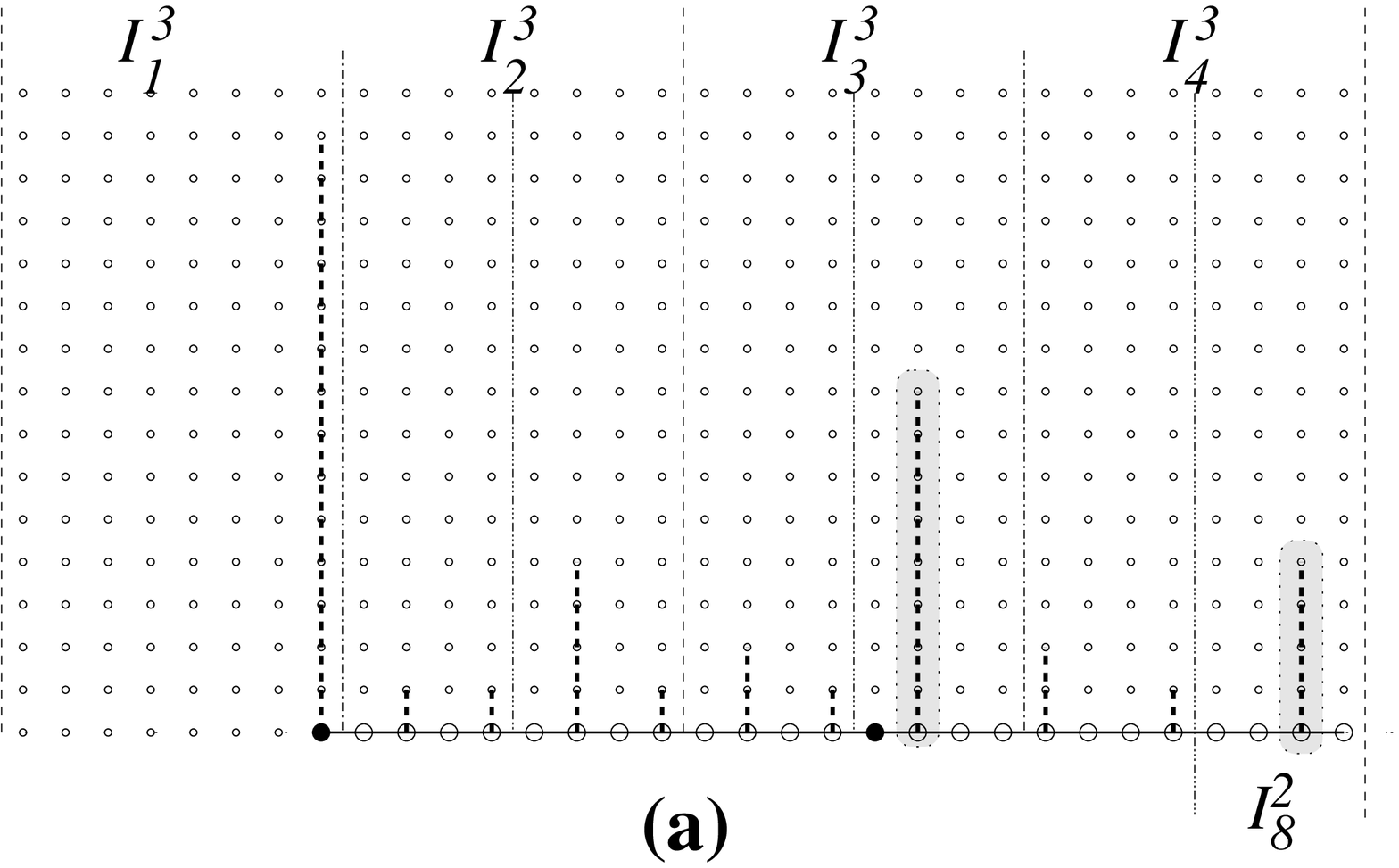}
\hfill
\includegraphics[width=0.4\textwidth]{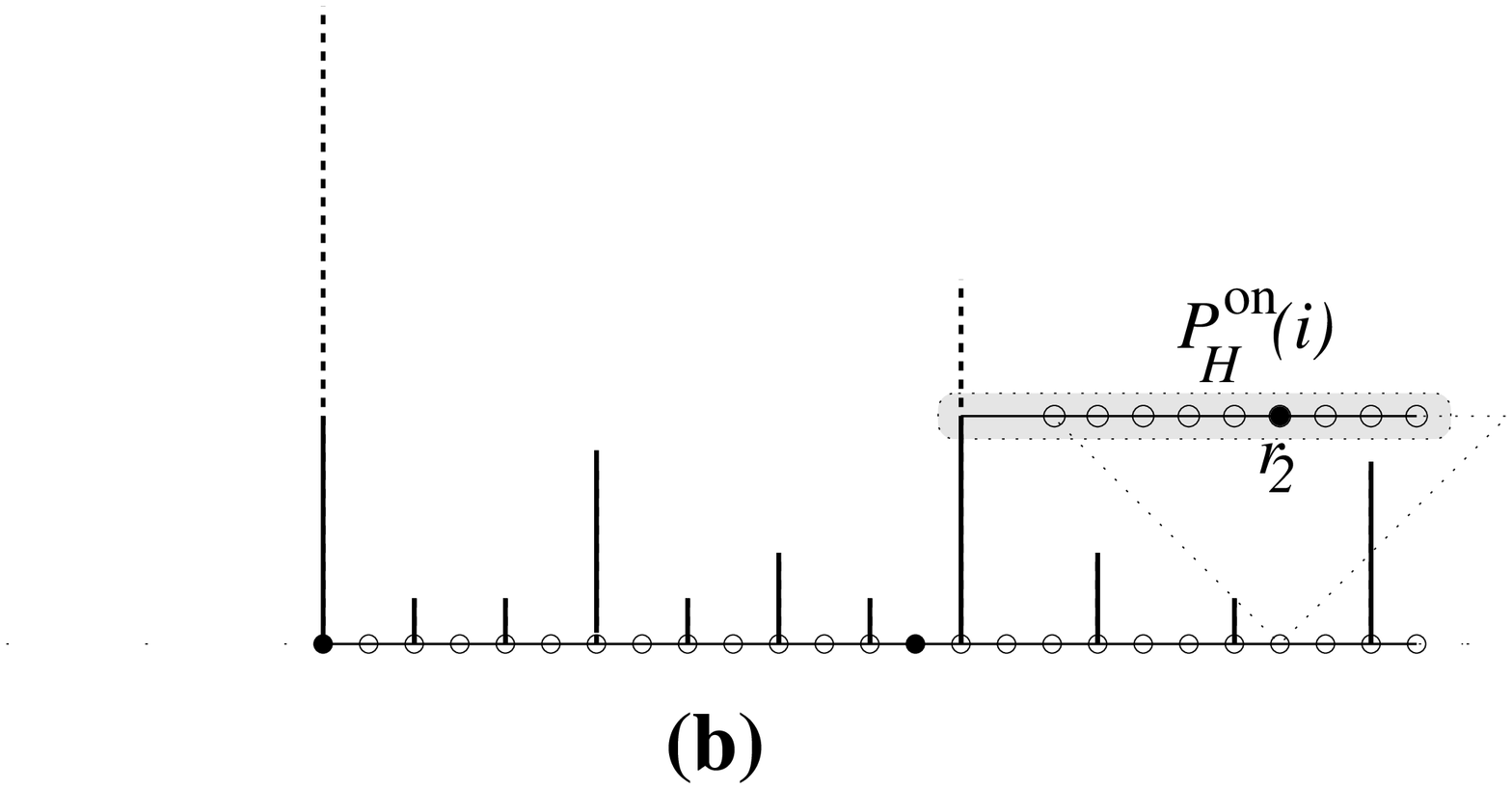}
\hfill
\includegraphics[width=0.4\textwidth]{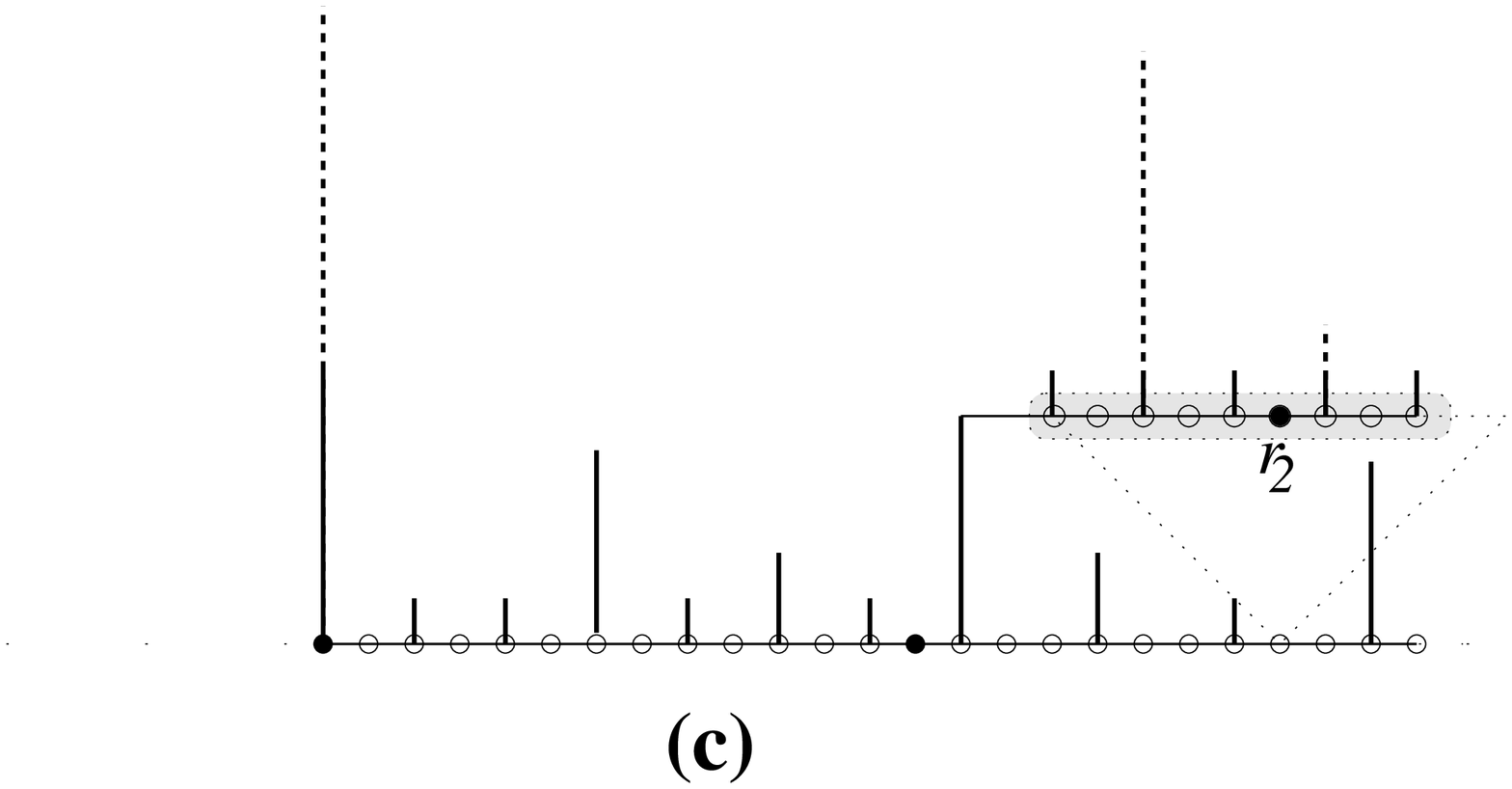}
\hfill
\includegraphics[width=0.4\textwidth]{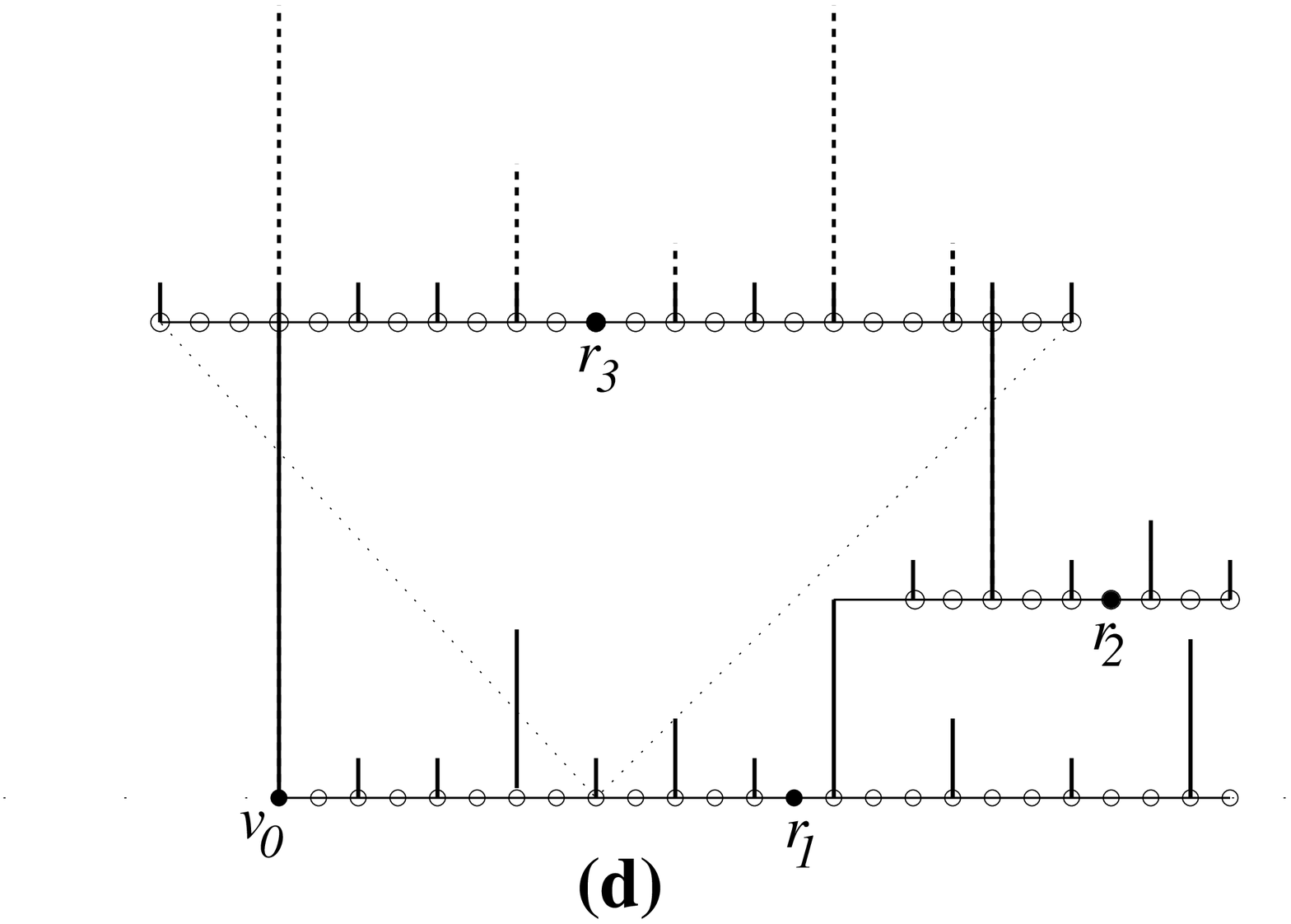}
\end{center}
\caption{\sf Example of execution of $\lineon$ on $\calR=\{v_0,r_1,r_2,r_3\}$.
The dashed segments show were $\lineon$ ``plans'' (it may changes its plans, when some requests arrives) to stores copies.
(a) $\lineon$ handles request $r_1$;
(b) delivery phase of $\lineon$ for $r_2$;
(c) storage phase of $\lineon$ for $r_2$;
(d) delivery and storage phases of $\lineon$ for $r_3$.
\label{figure:lineon}
 }
\end{figure*}

The pseudo code of $\lineon$
and an example for an execution of $\lineon$ are given
in Fig. \ref{figure: onalg} and Fig. \ref{figure:lineon}, respectively.
The solution constructed by $\lineon$ is denoted $\Fon=\Hon\cup\Aon$, where
$\Hon=\cup_{i=1}^{\NN}\PHon(i)$
represents the horizontal edges added in the delivery phases and
$\Aon=\{(\rep{v}{t},\rep{v}{t+1})\mid \rep{v}{t+1}\in\calC_{t+1} \mbox{ and } t=0,...,t_\NN\}$
represents the arcs added in the storage phase.
Before the main analysis, we make some easy to prove but crucial observations.
Recall that the notation of active (including $\stayactive$)
refer to the fact the nodes of some base replicas belong to some interval $I$ in the past.
Observations \ref{obser: complete onalg} and \ref{obser: I is active => N(I)cap Ct neq emptyset} state,
intuitively, that $\lineon$ leaves a copy in the {\em neighborhood} $\NI{I}$ of $I$
as long as $I$ is active.

\begin{observation}
{\sc (``Well defined'').}
If an interval $I\in\calI$ is $\stayactive$ at time $t$, then there exists a replica $\rep{v}{t}\in \calC_{t}\cup\Base[t]$ such that $v\in\NI{I}$. 
\label{obser: complete onalg}
\end{observation}
\def\ObserCompONALG{
\begin{proof}
Consider some interval $I\in\calI$ and a time $t$. If $I$ is $\stayactive$ at $t$,
then either $I[t]\cap\Base[t]\not=\emptyset$ (caused by a new request)
or $I[t]\cap\Base[t]=\emptyset$ and
$I$ is also $\stayactive$ at time $t-1$ (and
$\calC_{t}\cap \NI{I}[t]\not=\emptyset$);
hence, $(\Base[t]\cup \calC_{t})\cap \NI{I}[t]\not=\emptyset$.
The observation follows.
\end{proof}
\QED
} 
\ObserCompONALG

Moreover, a $\stayactive$ interval keeps a copy in its neighborhood longer.

\begin{observation}
{\sc (``An active interval has a near by copy'').}
If an interval $I$ is $\Active$ at time $t$, then, either (\rmnum{1}) there is some base replica in $I$'s
neighborhood at $t$ ($\Base\cap\NI{I}[t]\not=\emptyset$),
or (\rmnum{2}) at least one of the nodes of $\NI{I}$ stores a copy for time $t$ ($\NI{I}[t]\cap \calC_{t}\not=\emptyset$).
\label{obser: I is active => N(I)cap Ct neq emptyset}
\end{observation}
\def\ObserActiveNearByCopy{
\begin{proof}
Consider an interval $I\in\calI$ that is $\Active$ at time $t$.
If $\Base\cap\NI{I}[t]\not=\emptyset$, then the observation follows.
Assume that $\Base\cap\NI{I}[t]=\emptyset$.
Then, the fact that $I$ is active at $t$, but not contain any base replica at time $t$, implies also, that $I$ $\stayactive$ at time $t-1$.
Thus,  either (\rmnum{1}) $I$ commit at $t-1$ (at step (1)) which ``cause'' adding an additional replica to $\calC_t$ from $I$'s neighborhood;
or ($\rmnum{2}$) $I$ does not commit at $t-1$, since $\calC_t$ has, already, a replica from $I$'s neighborhood.
\end{proof}
\QED
} 
\ObserActiveNearByCopy

\begin{observation}
{\sc (``Bound from above on $|\Aon|$'').}
$|\Aon\setminus \calP_\calA[\rep{\rot}{0},\rep{\rot}{t_{\NN}}]|\leq |\COMMIT|$.
\label{obser: |Commit|=|Charge|=|Aon-0|}
\end{observation}
\def\ObserCommitAr{
\begin{proof}
Let
\commdouble\\\commdoubleend
$\Aon_{-\rot}=\Aon\setminus\calP_\calA[\rep{\rot}{0},\rep{\rot}{t_{\NN}}]$.
Now we prove that $|\Aon_{-\rot}|=|\COMMIT|$.
Every arc in $\Aon_{-\rot}$ (that add at step (S1.3)) corresponds to exactly one tuple $\langle I,t\rangle$ of an interval $I$ that commits at time $t$ (in step (S1.1));
and every interval commits at most once in each time $t$ that corresponds to exactly one additional arc in $\calA_{-\rot}$.
Thus, $|\Aon_{-\rot}|=|\COMMIT|$.
The observation follows.
\end{proof}
\QED
} 
\ObserCommitAr

\begin{figure*}
\fboxsep=0.2cm
\framebox[\textwidth]{
\begin{minipage}{0.95\textwidth}

~$\bullet$ $\Hon\leftarrow\emptyset$; $\Aon\leftarrow\emptyset$ and /* $\COMMIT\leftarrow\emptyset$; */

~$\bullet$ At time $t$ do:
    \begin{enumerate}

    \item If $\calR[t]\not=\emptyset$, then letting $\calR[t]=\{\rr_j,...,\rr_k\}$ and do:\\
    \textbf{\underline{``Delivery phase''}}$\vartriangleright$ deliver a copy to every base replica in $\Base[t]$.
    \begin{enumerate}

        \item ``Simulate'' $\Triangle$ and compute $\rhoT{j},...,\rhoT{k}$ and $\Base(j),...,\Base(k)$.

        \item For $i=j$ to $k$ do:
        \begin{enumerate}
            \item If $i=j$, then
            \begin{enumerate}
                \item chose a closest replica $\qon_j\in\calC_t$ to $\rr_j$;\\
                $\vartriangleright$ step (D1), where $d(\rr_j,\qon_j)=\min\{d(\qq,\rr_j)\mid \qq\in\calC_t\}$.
                \item $\PHon(j)\leftarrow\Base_\calH(j)\cup\calP_\calH[\qon_j,\rr_j]$ and $\PVon(j)\leftarrow\{ \rr \mid (\rr,\rr')\in\PHon(j) \}$.
            \end{enumerate}

            \item Otherwise,
            \begin{enumerate}
                \item chose a closest replica $\qon_i\in\calC_t\cup\bigcup_{l=j}^{i-1}\PVon(l)$ to $\rr_i$;\\

                $\vartriangleright$  step (D1), where $d(\rr_i,\qon_i)=\min\{d(\qq,\rr_j)\mid \qq\in\calC_t\cup\bigcup_{l=j}^{i-1}\PVon(l)\}$.

                \item $\PHon(i)\leftarrow\Base_\calH(i)\cup\calP_\calH[\qon_i,\rr_i]$ and $\PVon(i)\leftarrow\{ \rr \mid (\rr,\rr')\in\PHon(j) \}$.
            \end{enumerate}

            \item $\calHT[t]\leftarrow\calHT[t]\cup\PHon(i)$.
            $\vartriangleright$ step (D2); add the path between $\rr_i$ and $\qon_i$ and the base edges of $\Base_\calH(i)$ to the solution.
        \end{enumerate}

    \end{enumerate}

    \textbf{\underline{``Storage phase''}}
    \item $\calC_{t+1}\leftarrow\{\rep{\rot}{t+1}\}$;

    \item For each level $l=0$ to $\log m$ do:
    \label{onalg: main loop}
    \begin{enumerate}

        \item While there exists a level $l$ interval $I\in\calI$ that is $\stayactive$ at $t$ and $\NI{I}[t]\cap \calC_{t+1}=\emptyset$ do:
        \label{onalg: while loop}
        $\vartriangleright$ part of step (S1);

        \begin{enumerate}
            \item /* $\COMMIT\leftarrow\COMMIT\cup\{\langle I,t\rangle\}$ */
            \label{onalg: update Commit}
            $\vartriangleright$ part of step (S1.1);

            \item Select a replica  $\rr=\rep{v}{t}\in\Base[t]\cup \calC_{t}$ such that $v\in\NI{I}$.\\
            $\vartriangleright$ step (S1.2);
            \item $\calC_{t+1}\leftarrow \calC_{t+1}\cup\{\rr\}$. 
                        \label{onalg: update Ct+1}
            $\vartriangleright$ step (S1.3);

        \end{enumerate}
    \end{enumerate}

    \item $\Aon\leftarrow\Aon\cup\{(\rep{v}{t},\rep{v}{t+1})\mid\rep{v}{t+1}\in\calC_{t+1}\}$;~$\vartriangleright$ step (S1.3);\\
    $\vartriangleright$ Store a copy in $v$ for the succussive time (for time $t+1$), for every $(v,t+1)\in \calC_{t+1}$.\\
    $\vartriangleright$ For every $(v,t)\in V[t]\setminus \calC_{t+1}$ do: If $v$ keeps a copy, then delete the copy for time $t+1$.

    \end{enumerate}

\end{minipage}
}
\caption{\label{figure: onalg}
Algorithm $\lineon$.
Comments (between /* */) contains auxiliary actions for the analysis.}
\end{figure*}

\subsection{Analysis of $\blineon$}
\label{subsec: Analysis of online ALG}
We, actually, prove that
\begin{eqnarray*}
\frac{\cost(\lineon,\calR)}{\cost(\Triangle,\calR)}=O(\sqrt{\log n}),
\end{eqnarray*}
This implies the desired competitive ratio of $O(\sqrt{\log \nn})$ by Theorem \ref{thm: Triangle is a 3-approx}.
Proving a competitive ratio by comparing an online algorithm to an approximation algorithm
(rather then to the unknown adversary) may be a useful approach for other competitiveness proofs.
We first show, that the number of horizontal edges in $\Hon$ ({\em``delivery cost''}) is $O\left(\Delta\cdot\cost(\Triangle,\calR)\right)$.
Then, we show, that the the number of arcs in $\Aon$ ({\em``storage cost''}) is $O\left(\frac{\log \nn}{\Delta}\cdot\cost(\Triangle,\calR)\right)$.
Optimizing $\Delta$, we get a competitiveness of $O(\sqrt{\log \nn})$.

\paragraph{\bf Delivery cost analysis.}
For each request $\rr_i\in\calR$, the delivery phase (step (D2)) adds $\PHon(i)=\calP_\calH[\qon_i,\rr_i]\cup\Base_\calH(i)$
to the solution.
Define the {\em online} radius of $\rr_i$ as $\ron{i}=d(\qon_i,\rr_i)$.
Since $|\Base_\calH(i)|\leq 2\rhoT{i}$, it follows that,
\begin{eqnarray}
|\Hon|\leq \sum_{i=1}^{\NN} \left(\ron{i}+2\rhoT{i}\right).
\label{Ineq: |Hon| leq sum d(qon,p) + 2roff}
\end{eqnarray}
It remains to bound $\ron{i}$ as a function of $\rhoT{i}$ from above.
Intuitively, $\rhoT{i}$ includes the distance from some base replica $\qq_i=\rep{u_i}{s_i}\in\Base$ to
$\rr_i=\rep{v_i}{t_i}$.
That is, $\rhoT{i}$ includes the distance from $v_i$ to $u_i$ and the time difference between $s_i$ and $t_i$.
Restating Observation \ref{obser: I is active => N(I)cap Ct neq emptyset} somewhat differently (Claim \ref{claim: dist(v, C t+rho) leq 4 Delta rho} below),
we can use the distance
$|v_i-u_i|\leq\rhoT{i}$ and the time difference $t_i-s_i\leq\rhoT{i}$ for bounding $\ron{i}$.
That is, we show the $\lineon$ has a copy at time $t_i$ (of $\rr_i$) at a distance at most $4\Delta\rhoT{i}$ from $u_i$ (of $\qq_i$).
Since, $|v_i,u_i|\leq \rhoT{i}$, $\lineon$ has a copy at distance at most $(4\Delta+1)\rhoT{i}$ from $v_i$ (of $\rr_i$).

\begin{figure*}
\begin{center}
\includegraphics[width=0.4\textwidth]{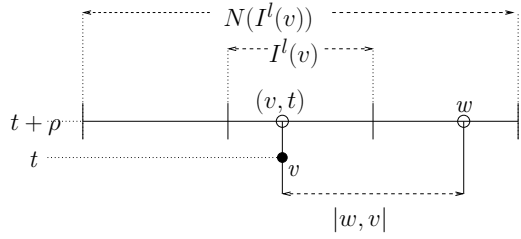}
\end{center}
\caption{\sf
\label{figure:AnaDevCostClaim}
Interval $\Interk{v}{l}$ is active at $t+\rho$, since $2^l\geq\rho$.
Therefore, there exists a replica $\qq=(w,t+\rho)\in\NI{\Interk{v}{l}}[t+\rho]\cap \calC_{t+\rho}$ and in addition,
$|w-v|\leq 2\Delta2^{l}$.
}
\end{figure*}

\begin{claim}
Consider some base replica $\rep{v}{t}\in\Base$ and some $\rho>0$, such that, $t+\rho\leq t_{\NN}$.
Then, there exists a replica $(w,t+\rho)\in\calC_{t+\rho}$ such that $|v-w|\leq 4\Delta\rho$ (Fig. \ref{figure:AnaDevCostClaim}).
\label{claim: dist(v, C t+rho) leq 4 Delta rho}
\end{claim}
\def\claimDevCostA{
\begin{proof}
Assume that $\rep{v}{t}\in\Base$.
Consider an integer $\rho>0$.
Let $l=\lceil \log \rho\rceil$.
Interval $\Interk{v}{l}$ is active at time $t+\rho$.
Thus, by Observation \ref{obser: I is active => N(I)cap Ct neq emptyset}, there exists some node in $\Interk{v}{l}$'s neighborhood that keep a copy for time $t+\rho$.
That is, a replica $\qq=(w,t+\rho)\in\NI{\Interk{v}{l}}[t+\rho]\cap \calC_{t+\rho}$ does exists.
The fact that $\qq\in\NI{\Interk{v}{l}}[t+\rho]$ implies that $w\in\NI{I}$, which implies that $|v-w|\leq2\cdot\Delta2^{l}$.
The claim follows, since $2\rho>2^{l}$.
\end{proof}
\QED
} 
\claimDevCostA

\begin{figure*}
\begin{center}
\includegraphics[width=0.4\textwidth]{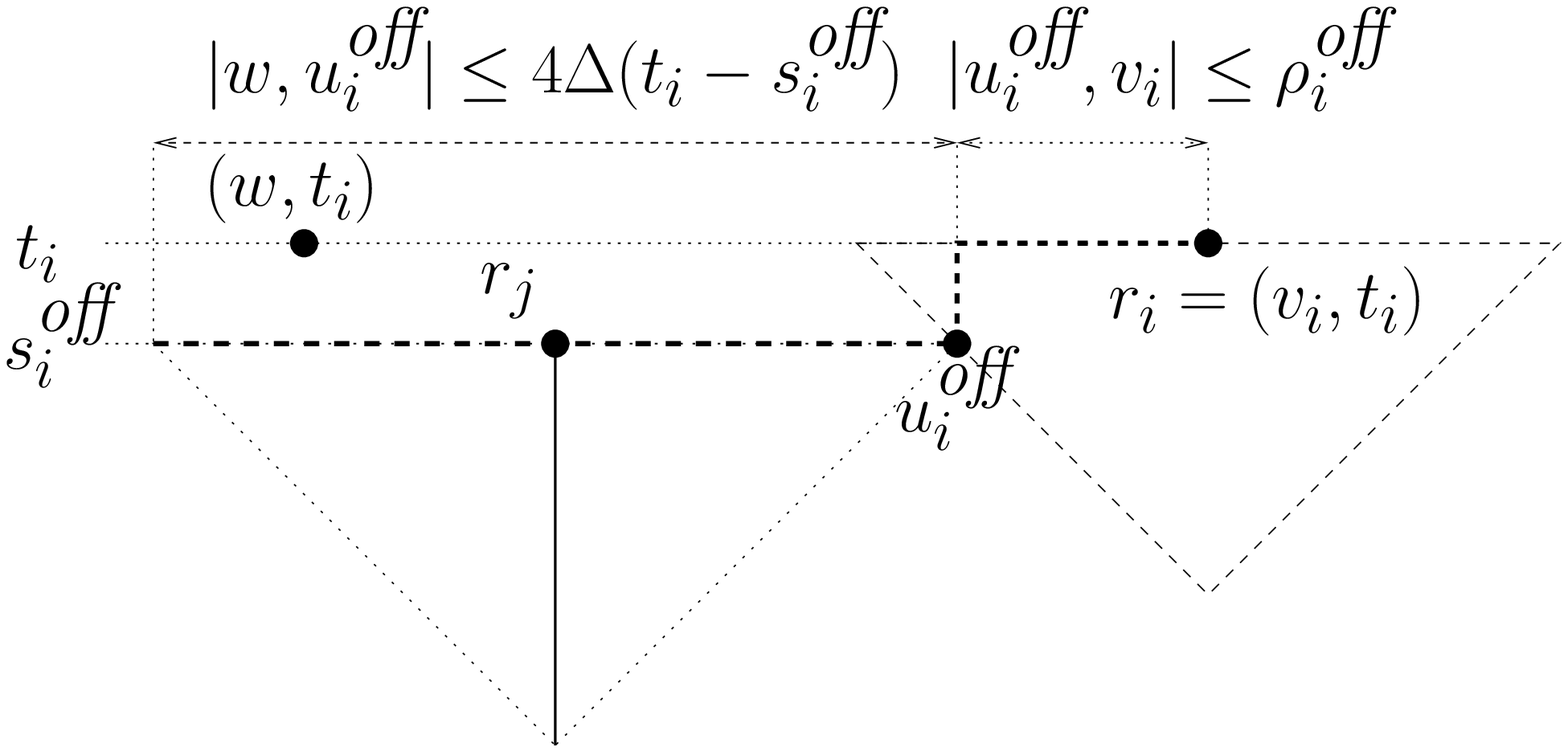}
\end{center}
\caption{\sf \label{figure:AnaDevCostLemma}
We have $|v_i,\uT_i|\leq\rhoT{i}$ and by Claim \ref{claim: dist(v, C t+rho) leq 4 Delta rho}, $|w,\uT_i|\leq 4\Delta\rhoT{i}$.
Thus, $\ron{i}\leq |v_i,w| \leq |v_i,\uT_i|+|\uT_i,w|\leq (4\Delta+1)\rhoT{i}$.
}
\end{figure*}

\begin{lem}
$\ron{i}\leq (4\Delta+1)\cdot\rhoT{i}$.
\label{lemma: delivery cost}
\end{lem}
\def\LemmaDevCost{
\proof
Recall that $\Triangle$ serves request $\rr_i=(v_i,t_i)$ from some base replica $\qT_i=(\uT_i,\sT_i)$ already include in the solution.
That $\qT_i$ may correspond to some earlier time. That is, $\sT_i\leq t_i$.
In the case that $\sT_i=t_i$, $\lineon$ can serve $\rr_i$ from $\qT_i$.
Hence, $\ron{i}\leq\rhoT{i}$.
In the more interesting case (see Fig. \ref{figure:AnaDevCostLemma}), $\sT_i<t_i$.
By Claim \ref{claim: dist(v, C t+rho) leq 4 Delta rho}
(substituting $v=\uT_i$, $t=\sT_i$, and $\rho= t_i-\sT_i\leq \rhoT{i}$),
there exists a replica $(w,t_i)\in\calC_{t_i}$
such that $|\uT_i,w|\leq 4\Delta\rhoT{i}$.
Recall that $|\uT,v_i|\leq d(\qT_i,\rr_i)=\rhoT{i}$.
Thus, by applying the triangle inequality, we get that,
$|v_i,w|\leq|w,\uT_i|+|\uT_i,v_i|\leq (4\Delta+1)\rhoT{i}$.
Hence, $\ron{i}\leq(4\Delta+1)\rhoT{i}$ as well.
\QED
} 
\LemmaDevCost

The following corollary follows from the above lemma, Inequality (\ref{Ineq: |Hon| leq sum d(qon,p) + 2roff}), and Theorem \ref{thm: Triangle is a 3-approx}.

\begin{corollary}
$|\Hon|\leq (4\Delta+3)\cdot|\opt|$.
\label{corollary: delivery cost}
\end{corollary}

\paragraph*{\bf\large Storage cost analysis.}
By Observation \ref{obser: |Commit|=|Charge|=|Aon-0|},
it remains to bound the size of $|\COMMIT|$ from above.
Let $\commit(I,t)=1$ if $\langle I,t\rangle\in\COMMIT$ (otherwise 0).
Hence, $|\COMMIT|=\sum_{I\in\calI}\sum_{t=0}^{\infty}\commit(I,t)$.
We begin by bounding the number of commitments in $\lineon$ made by level $l=0$ intervals.
\begin{observation}
$\sum_{I\in\{J\in\calI\mid \lfun{J}=0\}}\commit(I,t)\leq\big|\Base\big|.$
\label{obser: Base acounts level l=0 committments}
\end{observation}
\def\ObsCommitCountLevelZeroIntervals{
\begin{proof}
Consider some commitment $\langle I,t\rangle\in\COMMIT$, where interval $I$ is of level $\lfun{I}=0$.
Interval $I$ commit at time $t$ only if $I$ $\stayactive$ at $t$ (see step (S1) in $\lineon$).
This $\stayactive$ status at time $t$ occur only if there is base replica in $I$.
Moreover, the base replica must be at time $t$ since a base replica at $t$ cause an interval of level $l=0$ to be $\stayactive$ only at $t$.
Hence, each base replica causes at most one commitment at $t$ of one interval of level $l=0$.
Thus, $I$ is $\stayactive$ just at the times that $I$ has some base replicas.
\end{proof}
\QED
} 
\ObsCommitCountLevelZeroIntervals

The following is our main lemma;
\begin{lem}
$|\COMMIT|
\leq 3\big|\calAT_{}\big|+\frac{6\log n}{\Delta}\big|\calHT\big|+|\Base|
$.
\label{lem: storage cost < Hoff + Aoff}
\end{lem}
\noindent{\bf Proof sketch.}
The $|\Base|$ term in the statement of the lemma follows from Observation \ref{obser: Base acounts level l=0 committments} for level $l=0$ intervals.
The rest of the proof deals with commitments in intervals $I\in\calI$ whose level $\lfun{I}>0$.
We now group the commitments of each such an interval into {\em ``bins''}.
Later, we shall ``charge'' the commitments in each bin on certain costs of the offline algorithm $\Triangle$.

Consider some
level $l>0$ interval $I\in\calI$ and an input $\calR$.
We say that $I$ is a {\em committed-interval} if $I$ commits at least once in the execution of $\lineon$ on $\calR$.
For each committed-interval $I$ (of level $\lfun{I}>0$),
we define (almost) non-overlapping {\em``sessions''}
(one session may end at the same time the next session starts;
hence, two consecutive sessions may overlap on their boundaries).
The first session of
$I$ does {\em not} contain any commitments (and is termed an {\em uncommitted-session}); it begins at time $0$ and ends at the first time that $I$ contains some base replica.
Every other session (of $I$) contains at least one commitment (and is termed a {\em committed-session}).

Each commitment (in $\lineon$) of $I$ belongs to some committed session.
Given a commitment $\langle I,t\rangle\in\COMMIT$ that $I$ makes at time $t$,
let us identify $\langle I,t\rangle$'s session. 
Let $t^{-}< t$ be the last time (before $t$) there was a base replica in $I$.
Similarly, let $\tplus{}> t$ be the next time (after $t$) there will be a base replica in $I$
(if such a time does exist; otherwise, $\tplus{}=\infty$).
The session of commitment $\langle I,t\rangle$ starts at $\tminus{}$ and ends at $\tplus{}$.
Similarly, when talking about the $i$'s session of interval $I$, we say that the session starts at $\tminus{i}(I)$
and ends at $\tplus{i}(I)$.
When $I$ is clear from the context, we may omit $(I)$.
A bin is a couple $(I,i)$ of a commitment-interval and the $i$th commitment-session of $I$.
Clearly, we assigned all the commitments (of level $l>0$ intervals) into bins.

\begin{observation}
The bins do not overlap (except, perhaps, on their boundaries).
\label{Obser:MCD: bin do not overlap}
\end{observation}
\def\AppMCDObsrBinDoNotOverlap{
\proof
The sessions boundaries are times when $I$ has base replicas.
At those times, $I$ does not commit, since only level $l=0$ intervals may commit when they have a base replica
(if there exists a base replica in $I$ at time $t$,
then $I$ must contains some level $l=0$ interval $J^0\subseteq I$ that is $\stayactive$ at $t$;
recall that $\lineon$ deals (in the storage phase) with $J^0\in\calI$ of level $l=0$ before dealing with $I$;
one case is that $J^0$ commits (see (S1.1)) in $\lineon$ and store a copy (see (S1.2) and (S1.3)) in the neighborhood of $J^0$ and, hence, of $I$;
even $J^0$ {\em  may} not need to commit, if the solution of $\lineon$ already has a copy in the neighborhood of $J^0$ and, hence, of $I$;
thus, $I$ does not need to commit (see (S1)) in $\lineon$).

Therefore, there is no overlap between the sessions, except the ending and the starting times.
That is, $\tminus{0}\leq\tplus{0}\leq\tminus{1}<\tplus{1}\leq,...,\leq\tminus{i'}<\tplus{i'}$,
($i'$ is the number of bins that $I$ has).
\QED
} 
\AppMCDObsrBinDoNotOverlap

Let us now point at costs of algorithm $\Triangle$ on which we shall ``charge'' the set of commitments $\COMMIT(I,i)$ in bin $(I,i)$.
We now consider only a bin $(I,i)$ whose committed session is not the last.
Note that the bin corresponds to a rectangle of $|I|$ by $t_i^+ - t_i^-$ replicas. Expand the bin by $|I|$ replicas left and $|I|$ replicas right, if such exist (to $I$'s neighborhood $\NI{I}$).
This yields the {\em payer} of bin $(I,i)$; that is the payer is a rectangle subgraph of $|\NI{I}|$
by $t_i^+ - t_i^-$ replicas.
We point at specific costs $\Triangle$ had in this payer.

Recall that every non last session of $I$ ends with a base replica in $I$.
Let $\rep{v}{\tplus{i}}\in\Base\cap I[\tplus{i}]$ be some base replica in $I$ at the ending time of the session.
The solution of $\Triangle$ must contain a route ($\Triangle$ route) that starts at the root and reaches $\rep{v}{\tplus{i}}$
by the definition of a base replica.
For the charging, we use {\em some} (detailed below) of the edges in the intersection of the $\Triangle$ route and the payer rectangle.

\begin{figure*}
\begin{center}
\includegraphics[width=0.5\textwidth]{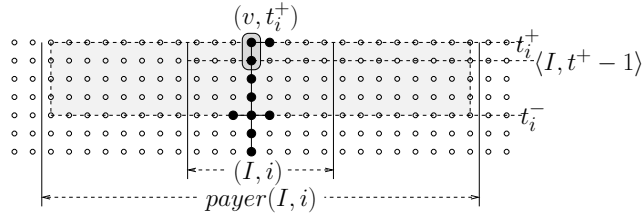}
\end{center}
\caption{\sf
\label{figure:BE}
$\Triangle$ route entrance from below (EB case);
Interval $I$ commits at $t$ and there exists an offline arc in $I$'s neighborhood from time $t$ to time  $t+1$. }
\end{figure*}

The easiest case is that the $\Triangle$ route enters the payer at the payer's bottom ($t_i^-$) and stays in the payer until $t_i^+$ (see Fig. \ref{figure:BE}).
In this case ({\bf EB}, for Entrance from Below), each time ($t_i^-  < t < t_i^+$) there is a commitment in the bin,
there is also an arc $a_t$ in the $\Triangle$ route (from time $t$ to time $t+1$).
We charge that commitment on that arc $a_t$.
Intuitively, the same arc $a_t$  may be charged also for one bin on the left of $(I,i)$ and one bin on its right, since the payer rectangles are 3 times wider than the bins.
Note that arc $a_t$ may also belong to additional $O(\log \nn)$ payers (of bins of intervals that contain $I$ or are contained in $I$). The crucial point is that
$a_t$ is {\em not} charged for those additional bins.
That is, we claim
that there are no commitments for those other bins. Intuitively, $\lineon$ was designed such that if $I$ commits at time $t$, $\lineon$ also stores a copy in $I$'s neighborhood for time $t+1$.
Hence, an interval $J$ whose neighborhood contains the neighborhood of $I$, does not need to commit (see the decision when not commit in (S1) in $\lineon$).
Thus, an arc of the $\Triangle$ route is charged only by 3 commitments at most
(this also proven formally later in Claim \ref{claim: an arc is assignd to 3 session}).

\begin{figure*}
\begin{center}
\includegraphics[width=0.45\textwidth]{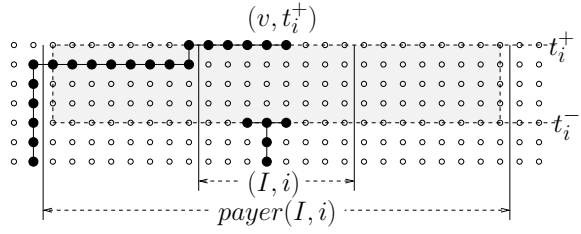}
\end{center}
\caption{\sf
\label{figure:SE}
$\Triangle$ route entrance from the left side (SE case) of the payer, that ``crosses'' the left neighbor of $I$;
thus, there exists at least $|I|$ horizontal edges in the intersection between the payer and $\Triangle$'s route.
}
\end{figure*}

The remaining case ({\bf SE}, for Side Entrance) is that
the $\Triangle$ route enters the payer from either the left or the right side of the payer.
(That is, $\Triangle$ delivers a copy from some other node $u$ outside $I$'s neighborhood,
rather than stores copies at $I$'s neighborhood from some earlier time,
See Fig. \ref{figure:SE}).
Therefore, the route must ``cross'' either the left neighbor interval of $I$ or the right neighbor interval in that payer.
Thus, there exists at least $|I|=\Delta2^{\lfun{I}}$ horizontal edges in the intersection between the payer ($\payer(I,i)$), of $(I,i)$  and
the $\Triangle$ route.

On the other hand, the number of commitments in bin $(I,i)$ is $2^{\lfun{I}}$ at most.
(To commit, an interval must be active; to be active, it needs a base replica in the last $2^{\lfun{i}}$ times;
a new base replica would end the session.)
That is, we charged the payer $\Delta$ times more horizontal edges than there are commitments in the bin.
On the other hand, each horizontal edge participates in $O(\log \nn)$ payers
(payers of 3 intervals at most in each level; and payers of 2 bins of each interval at most, since two consecutive sessions may intersect only at their boundaries).
This leads to the term $\frac{6\log \nn}{\Delta}$ before the $|\calHT|$ in the statement of the lemma.

For each interval $I$, it is left to account for commitments in $I$'s last session.
That is, we now handle the bin $(I,i')$ where $I$ has $i'$ commitment-sessions.
This session may not end with a base replica in $I$, so we cannot apply the argument above
that $\Triangle$ must have a route reaching a replica in $I$ at $\tplus{i'}$.
On the other hand, the first session of $I$ (the uncommitted-session) does end with a base replica in $I$, but has no commitments.
Intuitively, we use the payer of the first session of $I$ to pay for the commitments of the last session of $I$.
Specifically, in the first session, the $\Triangle$ route must enter the neighborhood of $I$ from the side;
(Note that the $\Triangle$ route still starts outside $I$; this because the origin $\rot$ who holds a copy, is not in $I$'s neighborhood; otherwise, $I$ would not have been a committed interval.)
Hence, we apply the argument of case SE above.
{\em (End of Proof sketch.) \QED }

\paragraph{Formal proof of the lemma.}
Extending the sketch into a some definitions omitted  from sketch.
Let us now start, give a formal definitions of the
aforementioned assignment of commitments to bins and the two charging assignments of offline horizontal edges and arcs.
Let $\BIN=\{(I,i) \mid I \mbox{ has at least } i \mbox{ bins}\}$.
For every bin $(I,i)\in\BIN$, let
$\combin(I,i)=\{\langle I,t\rangle \mid \tminus{i}\leq t\leq\tplus{i}\}$.

Let
$\payer(I,i)=\NI{I}[\tminus{i},\tplus{i}]$, if $(I,i)$ is not the last session of $I$, otherwise $\payer(I,i)=\NI{I}[\tminus{0},\tplus{0}]$.
Denote the charged set (of offline horizontal edges) for bin  $(I,i)$  by
$\horbin(I,i)= \calHT\cap\payer(I,i)$
and denote the charged set (of offline arcs) for bin $(I,i)$
by
\begin{eqnarray*}
\arcbin(I,i)= \{(\rep{v}{t},\rep{v}{t+1})\in\calAT\cap\payer(I,i)\mid \\\langle I,t\rangle\in\COMMIT\}.
\end{eqnarray*}

In addition to the above definitions, the following claim shows, formally,  that each offline arc is charged for 3 bins at most.
\begin{claim}
For every arc $a\in\calAT$,
\begin{center}
$|\{(I,i) \mid
a\in\arcbin(I,i)\}| \leq 3$.
\end{center}
\label{claim: an arc is assignd to 3 session}
\end{claim}
\proof
Denote the set in the statement of the claim by
$\Amin(a)=\{(I,i) \mid a\in\arcbin(I,i)\}$.
Consider an arc $a_{v,t}=((v,t),(v,t+1))\in\calAT$.
Recall that, $a_{v,t}\in\payer(I,i)$, if $v\in\NI{I}$ and $t\in[\tminus{i}(I),\tplus{i}(I)$.
Thus,
\begin{eqnarray*}
\Amin(a_{v,t})&=&
   \{\langle I,t\rangle\in\COMMIT \mid v\in\NLI{I}\} \cup\\
&& \{\langle I,t\rangle\in\COMMIT \mid v\in\NRI{I}\} \cup\\
&& \{\langle I,t\rangle\in\COMMIT \mid v\in I\},
\end{eqnarray*}
as $\NI{I}=\NLI{I}\cup I\cup \NRI{I}$.
We first analyze for the set corresponding to $v\in I$ rather than $v\in\NI{I}$.
We show that
\begin{equation}
|\{\langle I,t\rangle\in\COMMIT \mid v\in I\}|\leq 1.
\label{ineq: <I,t> in assign v in I}
\end{equation}
That is, we prove that
$$
|\{\langle \Interk{v}{l},t\rangle\in\COMMIT \mid l=0,...,\log m\}|\leq 1.
$$
Assume that there exists a level $l^*$ such that $\langle \Interk{v}{l^*},t\rangle\in\COMMIT$.
Consider some $\ell<l^*$. Assume (by way of contradiction) that $\langle \Interk{v}{\ell},t\rangle\in\COMMIT$.
Thus, in step {\bf(3)} of $\lineon$, some replica $\rep{u}{t+1}$ of a node $u\in\NI{\Interk{v}{\ell}}$ is added to $\calC_{t+1}$.
Thus, when $\lineon$ consider the $l^*$th iteration at time $t$,
the neighborhood of $\Interk{v}{l^*}$ at $t$, contains some replica (specifically, $\rep{u}{t+1}$) that belongs to $\calC_{t+1}$ (since $u\in\NI{\Interk{v}{\ell}}\subseteq\NI{\Interk{v}{l^*}}$).
Thus, $\langle \Interk{v}{l^*},t\rangle\not\in\COMMIT$.
This contradict the assumption that $\langle \Interk{v}{l^*},t\rangle\in\COMMIT$.
Now, consider some $l>l^*$.
The condition in step {\bf(1)} of $\lineon$,
implies that $\langle \Interk{v}{l},t\rangle\not\in\COMMIT$, since
$\NI{\Interk{v}{l^*}}\subseteq\NI{\Interk{v}{l}}$.
Hence, Ineq. (\ref{ineq: <I,t> in assign v in I}) holds.

\begin{figure}
\begin{center}
\includegraphics[width=0.4\textwidth]{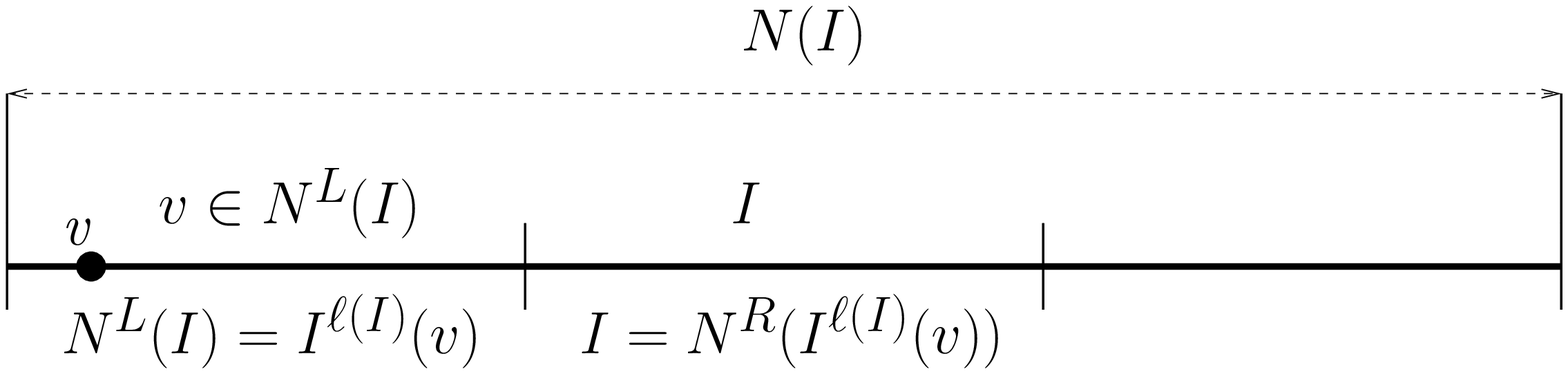}
\end{center}
\caption{\sf \label{figure:Neigh3}
$v\in\NLI{I}$, thus $\Interk{v}{\lfun{I}}=\NLI{I}$ and $\NRI{\Interk{v}{\lfun{I}}}={I}$.
}
\end{figure}

\begin{figure}
\begin{center}
\includegraphics[width=0.4\textwidth]{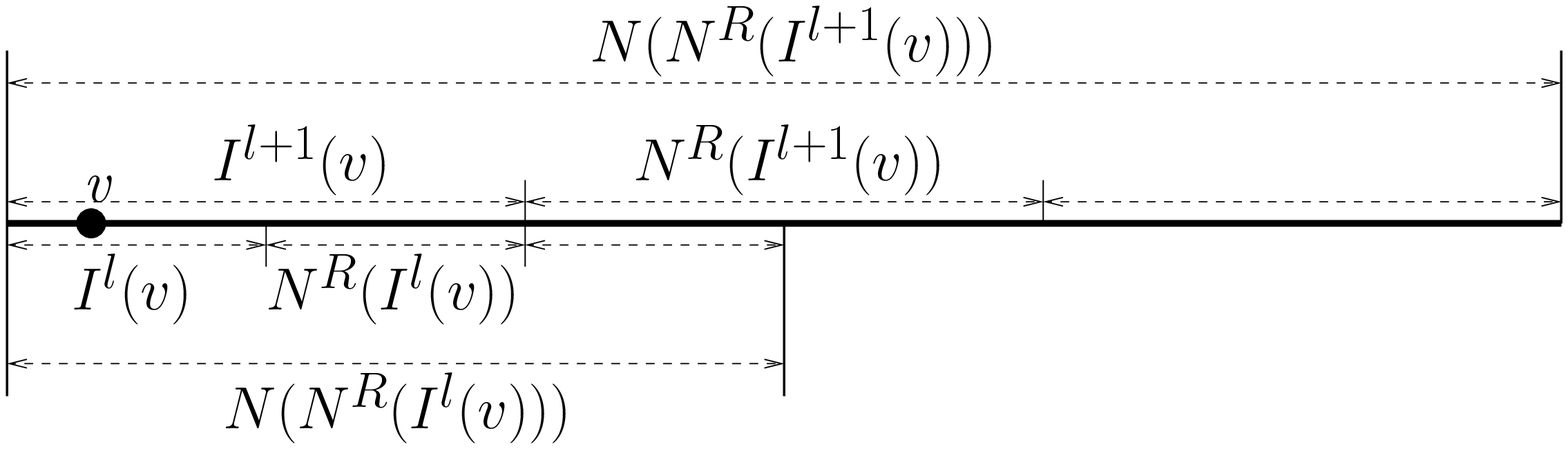}
\end{center}
\caption{\sf \label{figure:Neigh2}
We can see that $\NI{\NRI{\Interk{v}{l}}}\subseteq\NI{\NRI{\Interk{v}{l+1}}}$.
Thus, $\NI{\NRI{\Interk{v}{l'}}}\subseteq\NI{\NRI{\Interk{v}{l''}}}$, for every $l',l''\in\{0,...,\log m\}$ such that $l'\leq l''$.
}
\end{figure}

To prove that the claim holds, it is still left to prove similar inequalities for the set of left neighbors (of intervals that includes $v$) and for the set of right neighbors. First, let us show that,
\begin{equation}
|\{\langle I,t\rangle\in\COMMIT \mid v\in {\NLI{I}}\}|\leq 1.
\label{ineq: <I,t> in assign v in NLI(I)}
\end{equation}
We prove in fact, something equivalent. That is, we prove that
$
|\{\langle \NRI{\Interk{v}{l}},t\rangle\in\COMMIT \mid 
l=0,...,\log m\}|\leq 1$ (see Fig. \ref{figure:Neigh3}).
The proof is very similar to that of Ineq. (\ref{ineq: <I,t> in assign v in I}).
Assume that there exists a level $l^*$ such that
$\langle\NRI{\Interk{v}{l^*}},t\rangle\in\COMMIT$.
For every $l<l^*$, we have
$\NRI{\Interk{v}{l}}\subseteq\NRI{\Interk{v}{l^*}}$, while
for every $l>l^*$, we have $\NRI{\Interk{v}{l^*}}\subseteq\NRI{\Interk{v}{l}}$ (see Fig. \ref{figure:Neigh2}).
Because of the condition in step {\bf (1)} of $\lineon$,
we have that
$\langle \NRI{\Interk{v}{l}},t\rangle\not\in\COMMIT$, for every $l\in\{0,...,\log m\}\setminus\{l^*\}$.
Hence Ineq. (\ref{ineq: <I,t> in assign v in NLI(I)}) holds.
Similar arguments prove that
$
|\{\langle I,t\rangle\in\COMMIT \mid v\in {\NRI{I}}\}|\leq 1.
$
The claim follows by combining this together with inequalities (\ref{ineq: <I,t> in assign v in I}) and (\ref{ineq: <I,t> in assign v in NLI(I)}).
(claim \ref{claim: an arc is assignd to 3 session})
\QED

Let us now restate formally (but in a very formal condensed way) the claims defined informally in the sketch.
First, we bound the number of bins charging an horizontal edge.
As sketch above, each offline edge is charged for $6\log \nn$ bins at most.
Thus,
\begin{equation}
\sum_{(I,i)\in\BIN}|\horbin(I,i)|\leq 6\log \nn |\calHT|.
\label{ineq: sum horbin leq 3|Aoff|}
\end{equation}
At the same time Claim \ref{claim: an arc is assignd to 3 session} yields a bound on the number of bins charging an arc.
\begin{equation}
\sum_{(I,i)\in\BIN}(|\arcbin(I,i)|\leq 3|\calAT|.
\label{ineq: sum arcbin leq 3|Aoff|}
\end{equation}

It is left to count the number of edges and arcs assigned to each bin.
In case EB (the $\Triangle$ route enter the payer of bin $(I,i)$ from below), $|\combin(I,i)| \leq |\arcbin(I,i)|$.
In case SE, $\Delta|\combin(I,i)| \leq |\horbin(I,i)|$.
Thus, the edges and arcs assigned to bin $(I,i)$ obey
\begin{eqnarray}
|\combin(I,i)| \leq |\arcbin(I,i)|+\frac{1}{\Delta}|\horbin(I,i)|.
\end{eqnarray}
By Observation \ref{obser: Base acounts level l=0 committments},
\begin{equation}
|\COMMIT|\leq \sum_{(I,i)\in\BIN}(|\arcbin(I,i)|+\frac{1}{\Delta}|\horbin(I,i)|)+|\Base|.
\label{ineq: |COMMIT| leq sum (I,i) in session arcbin+Delta horbin}
\end{equation}
Now combine inequality (\ref{ineq: |COMMIT| leq sum (I,i) in session arcbin+Delta horbin}) with
inequalities
(\ref{ineq: sum horbin leq 3|Aoff|}) and (\ref{ineq: sum arcbin leq 3|Aoff|}).
Lemma \ref{lem: storage cost < Hoff + Aoff} follows.
\QED 

We now optimize a tradeoff between the storage coast and the delivery cost of $\lineon$.
On the one hand, Lemma \ref{lem: storage cost < Hoff + Aoff} shows that a large $\Delta$
reduces the number of commitments.
By Observation \ref{obser: |Commit|=|Charge|=|Aon-0|}, this means a large $\Delta$ reduces the storage cost of $\lineon$.
On the other hand, corollary \ref{corollary: delivery cost} shows that a {\em small} $\Delta$ reduces the delivery cost.
To balance this tradeoff, we need to ``manipulate'' Lemma \ref{lem: storage cost < Hoff + Aoff}
somewhat, since it uses variables that are different than those used in corollary \ref{corollary: delivery cost}.
We use the following observation
(1) $|\calP_\calA[\rep{\rot}{0},\rep{\rot}{t_{\NN}}]|\leq |\opt|\leq \cost(\Triangle,\calR)$; (2) $|\calAT|+|\calHT|=\cost(\Triangle,\calR)$; and (3) $|\Base{}|\leq \cost(\Triangle,\calR)$.
Substituting the above (1)--(3) in Observation \ref{obser: |Commit|=|Charge|=|Aon-0|} and Lemma \ref{lem: storage cost < Hoff + Aoff},
\begin{equation}
|\Aon|\leq \left(5+\frac{3\log \nn}{\Delta}\right)\cdot\cost(\Triangle,\calR).
\label{ineq: |Aon| leq 5 frac 3 Aoff + 6log n}
\end{equation}
To optimize the tradeoff, fix $\Delta=\sqrt{10\log \nn}$. Corollary \ref{corollary: delivery cost}, and inequality (\ref{ineq: |Aon| leq 5 frac 3 Aoff + 6log n}) imply that
$\cost(\lineon,\\\calR)=|\Aon|+|\Hon| \leq (8+\sqrt{10\log \nn})\cdot\cost(\Triangle,\calR)$.
Thus, by Theorem \ref{thm: Triangle is a 3-approx}, the following holds.

\begin{thm}
$\lineon$ is $O(\sqrt{\log \nn})$-competitive for $\MCD$ on the undirected line network.
\label{thm: onalg is sqrt(log NetSize) competitive}
\end{thm}

\section{Optimal online algorithm for {\large\bf\em SRSA}}
\label{sec: opt StRSA}

\newcommand{\ff}[0]{f}
\newcommand{\nguess}[0]{n\mbox{-}guess}
\newcommand{\Mguess}[0]{M\mbox{-}guess}
\newcommand{\tetration}[1]{2^{2^{2#1}}}
\newcommand{\costm}[0]{\widetilde{\cost}}
\newcommand{\FRSAn}[0]{\calF^{\mbox{\sc Srsa}}_{\nn}}
\newcommand{\FRSAnk}[1]{\calF^{\mbox{\sc Srsa}}_{\nn_{#1}}}
\newcommand{\FRSAmn}[0]{\calF^{\mbox{\sc Srsa}}_{\MM,\nn,p}}
\newcommand{\FRSAQ}[0]{\calF^{\mbox{\sc Srsa}}(\calQ)}
\newcommand{\FRSA}[0]{\calF^{\mbox{\sc Srsa}}}

Let us now transform $\lineon$ into an optimal algorithm for the online problem of $\StRSA$ \cite{berman}.
Note that without such a transformation, our solution for $\MCD$ (Section \ref{sec:MCD opt as a function of net size}) does not yet solve $\StRSA$.
In $\MCD$,  the $X$ coordinate of every request (in the set $\calR$) is taken from a known set of size $n$ (the network nodes $\{1,2,3, ... , n\}$).
On the other hand, in $\StRSA$, the $X$ coordinate of a {\em point} is arbitrary.%

The immediate idea how to bridge this problem is problematic. Intuitively, it looks as if it is enough just to translate the $X$ coordinates of points of $\StRSA$ into network nodes of $\MCD$. One problem in such an idea would be that in $\MCD$, the number of network nodes is known in advance, while the number of points in $\StRSA$ is not.

A more serious problem is somewhat more delicate. Intuitively, for $\lineon$ to work correctly,
the translation must maintain the proportion of the distances. That is, assume that some two points are very close to each other while some two other points are very far from each other. The first two points must be translated to network nodes that are close to each other, while the latter two points must be translated to
network nodes that are far from each other.  The competitive ratio of $\lineon$ on such an input would have been bad, since it would have depended on this proportion.

To overcome these problems, we first ``assume them away''. Then, we make a series of
modifications that remove the assumptions.
First, assume that we know in advanced a ``good'' guess $\nn$ on the number $\NN$ of points.
(Here, $n$ is a ``good'' guess if $\sqrt[4]{\nn}\leq \NN\leq \nn$;
intuitively, this ensures that $\sqrt{\log \nn}=\Theta(\sqrt{\log \NN})$; recall that
$O(\sqrt{\log \nn})$ is the upper bound we established for $\MCD$ and $O(\sqrt{\log \NN})$ is the upper bound are shooting for in this section for $\SRSA$.)
Also, assume that we know in advanced a ``good'' guess $\MM$ on $\xmaxQ=\max\{x_i\mid(x_i,y_i)\in\calQ\}$
the largest $X$ coordinate of any point.
(Specifically, here the guess $\MM$ is ``good'' if $\MM/2\leq\xmaxQ\leq \MM$;
intuitively, we pay $O(\MM)$ and $\opt$ pays $\Omega(\xmaxQ)$.)
Given those assumptions, we define a network (of $\MCD$) with $\nn$ nodes.
The length of a graph edge is thus, $\frac{\MM}{\nn}$ (less than  $\frac{2\xmaxQ}{\NN}$).
Another important assumption is not about our knowledge, but rather on the input itself.
That is, we assume that $\MM=\nn$.
(Though, later in $\MCD$, the length of an edge is ``normalized'' to 1.)
The details are left for the full paper.

\begin{thm}
Algorithm
$\onRSA$ is optimal and is $O(\sqrt{\log \NN})$-competitive.
\label{thm: onRSA}
\end{thm}

\section{Optimizing {\large\bf\em MCD} for a small number of requests}
\label{sec:Optimal-mcd-for-few-requests}

Algorithm $\lineon$ was optimal as the function of the network size (Theorem \ref{thm: onalg is sqrt(log NetSize) competitive}). This means that it may not be optimal in the case that the number of requests is much smaller than the network size. In this section, we use Theorem \ref{thm: onRSA} and algorithm $\onRSA$ to derive an improve algorithm for $\MCD$. This algorithm, $\lineonp$, is competitive optimal (for $\MCD$) for any number of requests.
Intuitively, we benefit from the fact that $\onRSA$ is optimal for any number of points (no notion of network size exists in $\StRSA$).

This requires the solution of some delicate point. Given an instance $\MCD^a$ of $\MCD$, we would have liked to just translate the set $\calR^a$ of $\MCD$ requests into a set $\calQ$ of $\StRSA$ points and
apply $\onRSA$ on them.
This may be a bit confusing, since $\onRSA$ performs by converting back to $\MCD$. Specifically,
recall that $\onRSA$ breaks $\calQ$ into several subsets, and translates back first the first subset $\calQ_1$ into an the requests set $\calR^b_1$ of a new instance $\MCD^b_1$ of $\MCD$. 
Then, $\onRSA$ invokes $\lineon$ on this new instance $\MCD^b_1$.
The delicate point is that $\MCD^b_1$ is different than $\MCD_a$.

In particular, the fact that $\calQ_1$ contains only {\em some} of the points of $\calR^a$, may cause $\onRSA$ to ``stretch'' their $X$ coordinates to fit them into the network of $\MCD_a$.
Going carefully over the manipulations performed by $\onRSA$
reveals that the solution of $\onRSA$ may not be a feasible solution
of $\MCD$ (even though it applied $\lineon$ plus some manipulations).
Intuitively, the solution of $\onRSA$  may ``store copies'' in places that are not grid vertices in the grid of $\MCD_a$. Thus the translation to a solution of $\MCD_1$ is not immediate.

Intuitively, to solve this problem, we translate a solution of $\onRSA$ to a solution of $\MCD_a$ in a way that is similar to the way we translated a solution of $\lineon$ to a solution of $\StRSA$. That is,  each request of $\MCD_a$ we move to a ``nearby'' point of $\onRSA$.
The details are left for the full paper.

\begin{thm}
Algorithm $\lineonp$ is optimal and it
\commdouble\\\commdoubleend
$O(\min\{\sqrt{\log \NN},\sqrt{\log \nn}\})$-competitive.
\end{thm}

\section{Randomized Lower Bound for the Line Network}
\label{sec:randomized_lb}

We obtain an $\Omega(\sqrt[3]{\log \nn})$ lower bound on the competitive
ratio of any {\em randomized} online algorithm for $\MCD$ on a line
network.
First, we describe a probability distribution $\calD$ on instances.
We show that the expected size of the solution returned by any
\emph{deterministic} algorithm executed on instances taken from
to $\calD$ is larger than the optimal offline solution by a factor of
$\Omega(\sqrt[3]{\log \nn})$.  The lower bound
then follows from Yao's min-max principle~\cite{Yao77}.
The details are left for the full paper.

\begin{thm}
The competitive ratio of any {\em randomized} online algorithm
for $\MCD$
on the line network is $\Omega(\sqrt[3]{\log n})$.
\end{thm}

\section*{Acknowledgment}
We would like to thank to Reuven Bar-Yehuda and Dror Rawitz for insights and helpful dissections.


\end{document}